%% file: main.tex
\documentclass[a4paper,fleqn]{cas-dc}

\input{preamble}

\begin{document}
\let\WriteBookmarks\relax
\renewcommand{\floatpagefraction}{0.85}
\renewcommand{\dblfloatpagefraction}{0.85}
\renewcommand{\textfraction}{0.05}
\renewcommand{\topfraction}{0.95}
\renewcommand{\dbltopfraction}{0.95}
\renewcommand{\bottomfraction}{0.95}
\setcounter{topnumber}{3}
\setcounter{dbltopnumber}{3}
\setcounter{totalnumber}{5}

\input{Content/Acronyms.tex}
\input{Content/macros_results.tex} 
\shorttitle{COHORT: Collaborative Orchestration for Hardening via Offensive Replay on Emulated Topologies}
\shortauthors{Frydman et~al.}
\title [mode = title]{COHORT: Collaborative Orchestration for Hardening via Offensive Replay on Emulated Topologies}                      
\author[1]{Chen Frydman}[orcid=0009-0006-1554-992X]

\author[2,1]{Aviram Zilberman}[orcid=0000-0002-8137-7314]
\cormark[1]

\author[1]{Rubin Krief}[orcid=0009-0000-9507-4368]

\author[1]{Abed Showgan}[orcid=0009-0006-6997-5043]

\author[3]{Andres Murillo}[orcid=0000-0001-6965-2283]

\author[3]{Sekiya Motoyoshi}[orcid=0009-0008-6542-2681]

\author[1]{Asaf Shabtai}[orcid=0000-0003-0630-4059]

\author[1]{Yuval Elovici}[orcid=0000-0002-9641-128X]

\author[1]{Rami Puzis}[orcid=0000-0002-7229-3899]

\affiliation[1]{organization={Ben-Gurion University of The Negev},
    city={Beer Sheva},
    country={Israel}}

\affiliation[2]{organization={Jerusalem College of Technology},
    city={Jerusalem},
    country={Israel}}

\affiliation[3]{op={}, organization={Fujitsu}}

\cortext[cor1]{Corresponding author}

\input{Content/Abstract}
\begin{keywords}
Network Emulation\sep Automatic Network Mitigation\sep Adversary Emulation\sep Cumulative Mitigations\sep Proactive Defense\sep Multi Agent System
\end{keywords}

\maketitle

\input{Content/Introduction}

\input{Content/Background_and_Related_Work}

\input{Content/Threat_Model}

\input{Content/The_Automatic_Mitigation_Framework}

\input{Content/Evaluation}

\input{Content/Discussion}

\input{Content/Conclusion}

\input{Content/Acknowledgements}


\bibliographystyle{cas-model2-names}

\bibliography{Bibliography/Bibliography}

\input{Content/Appendix}

\end{document}

%% file: preamble.tex
\usepackage[authoryear,longnamesfirst]{natbib}

\def\tsc#1{\csdef{#1}{\textsc{\lowercase{#1}}\xspace}}
\tsc{WGM}
\tsc{QE}
\tsc{EP}
\tsc{PMS}
\tsc{BEC}
\tsc{DE}

\usepackage{graphicx} 
\usepackage{booktabs}
\usepackage{amssymb}
\usepackage{tabularx}
\usepackage{graphicx}   

\providecommand{\Description}[1]{}

\usepackage{caption}
\usepackage{subcaption}

\usepackage{array}
\usepackage{multirow}
\usepackage{tabularx}
\usepackage{makecell}

\usepackage{xurl}      

\usepackage{listings}
\usepackage[most]{tcolorbox}
\usepackage{nameref}
\tcbuselibrary{listingsutf8}

\newtcolorbox{keyresult}{%
  colback=blue!8!white,
  colframe=blue!40!white,
  arc=2pt,
  boxrule=0.5pt,
  left=6pt, right=6pt, top=4pt, bottom=4pt,
  unbreakable,
  enhanced,
  before skip=6pt, after skip=6pt
}

\usepackage[utf8]{inputenc}  

\usepackage[nolist]{acronym}  

\usepackage[ruled,vlined]{algorithm2e}

\usepackage{enumitem}
\setlist{nosep}
\setlist[enumerate]{leftmargin=*}
\setlist[itemize]{leftmargin=*}


%% file: Content/Acronyms.tex
\begin{acronym}
  \acro{AI}{Artificial Intelligence}
  \acro{API}{Application Programming Interface}
  \acro{APT}{Advanced Persistent Threat}
  \acro{C2}{command-and-control}
  \acro{CoT}{Chain-of-Thought}
  \acro{CT}{Cyber Twin}
  \acro{CTI}{Cyber Threat Intelligence}
  \acro{DDoS}{Distributed Denial of Service}
  \acro{MSR}{mitigation success rate}
  \acro{EDR}{Endpoint Detection and Response}
  \acro{GNS3}{Graphical Network Simulator 3}
  \acro{ICS}{Industrial Control System}
  \acro{IDS}{Intrusion Detection System}
  \acro{IoC}{Indicators of Compromise}
  \acro{IM}{Incident Management}
  \acro{IR}{Incident Report}
  \acro{IPS}{Intrusion Prevention System}
  \acro{KG}{Knowledge Graph}
  \acro{LLM}{large language model}
  \acro{ME}{Mitigation Effectiveness}
  \acro{ML}{Machine Learning}
  \acro{NLP}{Natural Language Processing}
  \acro{OCE}{On Call Engineer}
  \acro{PE}{Prompt Engineering}
  \acro{RAG}{Retrieval-augmented generation}
  \acro{SIEM}{Security Information and Event Management}
  \acro{SOAR}{Security Orchestration Automation and Response}
  \acro{SOC}{Security Operations Center}
  \acro{TTP}{Tactics, Techniques, and Procedures}
  \acro{PT}{Penetration Testing}
  \acro{ASP}{Attack Success Parity}
\end{acronym}

%% file: Content/macros_results.tex

\newcommand{\msrSingle}{10.7}
\newcommand{\msrMultiNoCritic}{38.6}
\newcommand{\msrMultiCritic}{46.7}
\newcommand{\msrMultiCriticOverSingle}{4.4}
\newcommand{\connSingle}{92.0}
\newcommand{\connMultiNoCritic}{94.3}
\newcommand{\connMultiCritic}{88.0}
\newcommand{\plateauSingle}{28}
\newcommand{\plateauMultiNoCritic}{47}
\newcommand{\plateauMultiCritic}{52}
\newcommand{\msrTopoSmall}{48.9}
\newcommand{\msrTopoMedium}{48.5}
\newcommand{\msrTopoLarge}{42.6}
\newcommand{\plateauTopoSmall}{43}
\newcommand{\plateauTopoMedium}{45}
\newcommand{\plateauTopoLarge}{39}
\newcommand{\nTopoSmall}{188}
\newcommand{\nTopoMedium}{200}
\newcommand{\nTopoLarge}{195}
\newcommand{\nRunsSingle}{60}
\newcommand{\nRunsMultiNoCritic}{60}
\newcommand{\nRunsMultiCritic}{59}
\newcommand{\nRunsTotal}{179}
\newcommand{\nMitSingle}{600}
\newcommand{\nMitMultiNoCritic}{599}
\newcommand{\nMitMultiCritic}{583}
\newcommand{\nMitTotal}{1782}
\newcommand{\catHostHardeningN}{341}
\newcommand{\catHostHardeningMSR}{63.3}
\newcommand{\catAppControlN}{39}
\newcommand{\catAppControlMSR}{82.1}
\newcommand{\catHBIPSN}{13}
\newcommand{\catHBIPSMSR}{53.8}
\newcommand{\catNetTrafficFilterN}{139}
\newcommand{\catNetTrafficFilterMSR}{5.8}
\newcommand{\msrAttackDNSExfilCritic}{27}
\newcommand{\msrAttackLatMovCritic}{60}
\newcommand{\msrAttackRansomwareCritic}{51}
\newcommand{\msrAttackThiefCritic}{49}
\newcommand{\anovaTopoF}{0.99}
\newcommand{\anovaTopoDFOne}{2}
\newcommand{\anovaTopoDFTwo}{386.1}
\newcommand{\anovaTopoP}{0.37}
\newcommand{\anovaTopoPPairMin}{0.21}
\newcommand{\discoverySingleKOne}{50}
\newcommand{\discoverySingleKTwo}{70}
\newcommand{\discoverySinglePlateau}{78}
\newcommand{\discoveryMultiNoCriticKTwo}{75}
\newcommand{\discoveryMultiNoCriticPlateau}{95}
\newcommand{\discoveryMultiCriticPlateau}{98}
\newcommand{\discoveryResidualGap}{20}
\newcommand{\squidAssrBaseline}{72.22}
\newcommand{\squidAssrMitigated}{57.14}
\newcommand{\squidThresholdKB}{8}
\newcommand{\squidPayloadKB}{11.5}
\newcommand{\selfTestAssrBaseline}{0.83}
\newcommand{\selfTestAssrMitigated}{0.33}
\newcommand{\selfTestME}{0.5}

%% file: Content/Abstract.tex
\begin{abstract}
Mitigating an observed adversary in an enterprise network typically takes weeks of expert work: an analyst derives a mitigation tailored to that adversary, validates it without breaking production, and verifies it disrupts the specific attack.
The procedure relies on expert judgment and cannot safely be exercised against the production network. COHORT is the first end-to-end framework to automate this procedure for deployable mitigations.
A role-decomposed multi-agent LLM workflow proposes candidates, implements them as real device commands, and refines them through a critique loop, all on a high-fidelity GNS3 emulator running real vendor firmware (firewall, switch, router). Each candidate is evaluated by \emph{offensive replay}: re-executing the original adversary on the mitigated network for a paired comparison against the unmitigated baseline, rather than the reward-signal or expert-judgment proxies used in prior simulation, hybrid, and configuration-generation work. Two further checks complement replay: a connectivity-regression check (LAN ping and internet HTTP probe) rejects mitigations that disrupt legitimate LAN or internet connectivity, and a cumulative evaluation stacks approved mitigations onto a persistent state to surface compound effects. Across three topologies and four attack scenarios (ransomware, lateral movement, DNS exfiltration, data theft), \msrMultiCritic\% of generated mitigations both disrupt the attack and preserve connectivity under replay, \msrMultiCriticOverSingle$\times$ the rate of a single-agent baseline using the same model and tool access.
A demo video walking through the framework is available with our released artifacts (Appendix~\ref{appen:open-science}).
\end{abstract}

%% file: Content/Introduction.tex
\section{Introduction}\label{sec:introduction}

Given an adversary profile reconstructed from forensic analysis, threat-intelligence feeds, or red-team exercises, the defender must derive a deployable mitigation tailored to that adversary, validate it without breaking production while verifying that it disrupts the specific attack.
Two properties of this workflow drive its cost: it relies on expert knowledge, and it cannot safely be exercised on the production network, since experimentally adjusting firewall rules, routing policy, or host configuration risks breaking legitimate operation.
Existing reactive tooling (IDS/IPS, SIEM, SOAR) executes curated or playbook-driven responses for \emph{known} attack patterns; even AI-assisted SOAR operates within a maintained playbook set rather than synthesizing new controls~\cite{lyu2024survey,islam2019multivocal}.
Threat intelligence informs analysts on new attacks but does not generate per-incident defenses.
Breach-and-attack simulation (BAS) replays curated adversary techniques against the unmodified network to validate detection coverage rather than to test newly proposed mitigations.
The IBM Cost of a Data Breach report measures mean breach containment times of multiple weeks~\cite{ibm-data-breach-report}; per-incident mitigation derivation is a cost driver inside that window.

We present COHORT, the first end-to-end framework that autonomously generates and validates deployable mitigations.
To remove the expert-knowledge bottleneck, a multi-agent LLM workflow proposes, implements, and reviews each candidate mitigation as real device configuration commands.
To avoid production exposure during evaluation, the workflow operates on a high-fidelity emulated enterprise network running real vendor firmware (firewall, switch, router) in GNS3, with Caldera~\cite{applebaum2016caldera, chang2025characterizing} executing ATT\&CK-aligned adversary actions.
Validation rests on \emph{offensive replay}: from an established foothold, the same Caldera adversary is re-executed against the network with the candidate mitigation in place, and a \emph{judge} compares the per-step attack outcome to the unmitigated baseline.
Three further mechanisms shape generation and evaluation: an iterative implementation--critique loop catches configuration-text errors before replay; a connectivity-regression check (a LAN ping and an internet HTTP probe) rejects mitigations that disrupt legitimate LAN or internet connectivity; a cumulative evaluation stacks approved mitigations on a persistent network state to surface compound effects.

We target post-compromise enterprise networks against scripted, intelligence-equipped adversaries; adaptive adversaries are out of scope by construction (Section~\ref{sec:threat-model}).

Prior work covers parts of this workflow but not all of it: simulation and hybrid platforms (CybORG/CAGE~\cite{kiely2025exploring}, CyberBattleSim~\cite{msft:cyberbattlesim}, CSLE~\cite{hammar2022intrusion}) cannot test real configurations, configuration-generation systems (Stackelberg planning~\cite{speicher2019towards}, NetConfEval~\cite{wang2024netconfeval}) never confront an active adversary, emulation platforms (CyGIL~\cite{li2021cygil}) target attacker training rather than defender mitigation, concurrent LLM-based defense agents~\cite{saroui2025agentnirs} skip replay-based re-validation, and commercial BAS replays curated rather than generated mitigations. None covers the full workflow on a high-fidelity emulator using real device configuration commands (Section~\ref{sec:background-related}).

The mitigation workflow is decomposed into role-specialized 
LLM agents.
A \emph{suggester} proposes deployable defenses (firewall ACLs, routing-policy changes, host-hardening commands).
An \emph{implementer} translates a selected mitigation into device commands within a bounded per-cycle command budget.
A \emph{critic} reviews each implementation before replay; the implementation--critique loop runs to a fixed iteration limit before the mitigation is accepted or discarded.
Two further agents support evaluation: a \emph{judge} compares pre- and post-mitigation outcomes via Caldera replay, and a \emph{summarizer} produces a human-readable artifact.
The role-specialized design is contrasted against two ablations using the same model and tool access (Section~\ref{sec:single-agent}, Section~\ref{sec:experimental-conditions}): a single-agent baseline that consolidates these roles into one conversation, and a multi-agent variant with the iterative critique loop disabled.

We evaluate COHORT across small, medium, and large enterprise topologies emulated in GNS3, four attack scenarios (Section~\ref{sec:threat-model}), and \nMitTotal{} mitigation attempts, all driven by GPT-5.4~mini.
The multi-agent framework reaches an overall \ac{MSR} of \msrMultiCritic\%, where MSR is the rate of mitigations that both reduce attack progress (ME~$>$~0) and pass both connectivity probes (Section~\ref{sec:eval-metrics}).
This is \msrMultiCriticOverSingle$\times$ the single-agent baseline of \msrSingle\%, with the gap attributable to role specialization and the iterative critique loop (Section~\ref{sec:defense-success}).
Stacking approved mitigations on a persistent network state plateaus at \plateauMultiCritic\% cumulative attack-step reduction by the fourth defense (Section~\ref{sec:cumulative-effectiveness}).
We do not claim transfer to adaptive adversaries, Windows endpoints, model families beyond GPT-5.4 mini, or vendor families beyond FortiGate, Cisco IOS, and Open vSwitch (Section~\ref{sec:scope-generalization}); calibration against expert-derived mitigations is open (Section~\ref{dis:open-questions}).
All evaluation runs in an isolated emulator; production deployment of validated mitigations is a separate human-driven step that falls outside the framework.

\noindent\textbf{Contributions.}
\begin{itemize}
    \item \textbf{End-to-end framework that addresses both bottlenecks.} The first framework that autonomously generates, deploys, and validates deployable mitigations against the specific observed adversary, removing the expert-knowledge dependency and confining experimentation to a high-fidelity emulator with real device commands rather than the production network.

    \item \textbf{Multi-agent decomposition.} Lift from \msrSingle\% to \msrMultiCritic\% MSR over a matched single-agent baseline (same model, inputs, tool access). An ablation decomposes the lift: role specialization accounts for roughly three-quarters, the iterative critique loop for the remaining quarter.

    \item \textbf{Cumulative evaluation.} A non-regression acceptance rule accumulates approved mitigations on a persistent network state, exposing compound effects that per-mitigation evaluation cannot surface.

\end{itemize}

%% file: Content/Background_and_Related_Work.tex
\section{Background and Related Work}\label{sec:background-related}

We organize prior work along the six dimensions COHORT integrates: enterprise defense, automated configuration, multi-agent LLM reasoning, autonomous defense agents, the environments used to evaluate them, and commercial tooling. Each subsection ends by identifying the component of COHORT's workflow that the corresponding line of work omits.

\subsection{Enterprise Network Defense}
Modern enterprise defenses span firewalls, IDS/IPS, SIEM, SOAR, and EDR~\cite{lyu2024survey} under a fundamental security--availability tradeoff: a control that disrupts an attack but breaks legitimate services is not deployable.
Pre-deployment validation reasons over abstract infrastructure models, while security orchestration and AI-driven incident response automate the reactive side with curated playbooks~\cite{islam2019multivocal, hamadanian2023holistic}; neither synthesizes a new attack-specific mitigation, nor validates one by re-executing the observed adversary against the modified network.

\subsection{Automated Configuration and Network Management}
Automating network configuration is challenging due to ambiguity in intent, configuration conflicts, and unintended side effects~\cite{bringhenti2023automation}.
Recent LLM-based tools translate natural-language intents into device configurations: NetConfEval~\cite{wang2024netconfeval} benchmarks intent-to-config translation; GeNet~\cite{ifland2025genet} assists engineers with topology and configuration updates; LLM-NetCFG~\cite{lira2024large} targets zero-touch service management; and recent multi-vendor intent-driven generators~\cite{wang2026llm} extend the paradigm.
Recent surveys and validation tests across these systems assess intent fulfillment, syntactic correctness, and policy compliance; however, they do not evaluate how the generated configuration affects a re-executed adversary~\cite{hong2025comprehensive}.

\subsection{Large Language Models and Multi-Agent Reasoning in Cybersecurity}
\acp{LLM} enable contextual reasoning over heterogeneous security data~\cite{zhang2025llms} and have been applied to alert triage, threat explanation, and candidate remediation~\cite{wang2024shieldgpt, kaheh2023cyber}, but mostly in human-in-the-loop workflows and remain susceptible to hallucinations that fabricate threats or miss real ones, motivating explicit validation~\cite{sood2025paradigm}.
Multi-agent architectures decompose complex tasks into specialized roles~\cite{li2023camel, hong2023metagpt, an2024nissist, luo2026trustworthy}, with iterative refinement and coordinated reasoning across agents~\cite{du2024improving}.
Prior work focused on analysis and recommendation, scoring defenses through RL reward signals, expert judgment, or averages over independent attack runs. Isolating a particular mitigation's effect on the targeted adversary requires re-executing the same adversary against the just-mitigated network, a step that these methods omit.

\subsection{Autonomous Cyber Defense}
Autonomous defensive agents vary in their degree of autonomy and adaptability~\cite{stakhanova2007taxonomy}: \emph{scripted} agents execute fixed, predetermined action sequences with no runtime reasoning;
\emph{planning-based} agents dynamically construct action sequences by reasoning over a maintained model of current network state and goals;
\emph{reinforcement learning} (RL) agents learn policies through environmental interaction but remain opaque and difficult to transfer across environments~\cite{nguyen2021deep}; 
and \emph{large language model} (LLM)-based agents leverage pre-trained world knowledge and natural language reasoning to generate, critique, and explain actions, addressing the explainability and transferability limitations of RL-only approaches~\cite{castro2025large}.
RL-based agents learn mitigation policies in simulated enterprise environments~\cite{wang2024multiagent}, while LLM-based defenders (e.g., AgentNIRS~\cite{saroui2025agentnirs}, which generates firewall rules from NIDS alerts) and hybrid RL+LLM systems~\cite{loevenich2025design} drive containment in simulators, SOAR pipelines, and cyber ranges (Table~\ref{tab:acd-systems-2}).
On the offensive side, Caldera has been used to evaluate cyber deception against AI-based attackers~\cite{kouremetis2024mirage}, and pre-trained LLMs themselves have been deployed as autonomous attackers in cybersecurity RL environments such as NetSecGame, matching or exceeding purpose-trained RL baselines without per-environment training~\cite{rigaki2023out}.
Concurrent work begins to standardize evaluation for intrusion-response systems on public NIDS datasets~\cite{marchioro2025network}, but defensive agents otherwise operate in abstract simulators or execute reactive playbooks, and adversary emulation platforms execute attacks without proposing or validating mitigations (Table~\ref{tab:acd-systems-2}).

\subsection{Simulated and Emulated Network Environments for Adversarial Emulation}
Safe evaluation of autonomous cyber defense strategies requires isolated environments that can faithfully reproduce adversarial behavior without risk to production systems.
The literature spans three approaches with a fidelity--speed tradeoff.
\emph{Network simulation} uses abstract graph-based or Markov-decision-process representations~\cite{yamin2020cyber}; attack and defense execute as programmatic state transitions, so simulators cannot surface the side effects of real configuration changes.
\emph{Network emulation} instantiates real operating systems, protocols, and devices in virtualized hardware, with backends ranging from containers and VMs to full hardware virtualization booting vendor firmware images (e.g., GNS3~\cite{neumann2015book}); only the latter can evaluate deployable mitigations as they would behave on production hardware, building on the digital-twin paradigm of safe security testing on virtual replicas~\cite{eckhart2018towards}.
\emph{Hybrid} platforms couple a fast simulation surrogate with an emulated backend~\cite{molina2021network, janisch2023nasimemu, oesch2024towards}.

Beyond fidelity, we evaluate environments along four validation properties. The first two of these properties are operationalized in Section~\ref{sec:framework}: \emph{mitigation validation} (replay-based attack disruption) and \emph{connectivity regression testing} (verifying that legitimate LAN and internet reachability is preserved after a mitigation is deployed).
\emph{Deployable mitigations} require that defenses be expressed as real device configuration commands rather than abstract state transitions, so that what is validated is what would deploy.
\emph{Cumulative mitigations} compose multiple independently proposed defenses on a persistent network state, capturing synergistic or interfering effects that per-mitigation evaluation misses.

Across these environments, none provides all four validation properties: simulators cannot evaluate deployable mitigations, and even emulators that support offensive replay~\cite{kouremetis2024mirage, hammar2022intrusion} do not pair it with connectivity regression testing or cumulative mitigation evaluation.
Our framework addresses all four gaps by integrating Caldera adversary emulation with a GNS3-based emulated enterprise network that boots real vendor device images, enabling attack replay after mitigation deployment, automatic connectivity validation, and cumulative evaluation of stacked defenses (Table~\ref{tab:environments}).

\subsection{Commercial Adversary Emulation and Continuous Validation}
Industry tooling spans three overlapping categories: Breach-and-Attack Simulation (AttackIQ~\cite{attackiq}, Cymulate~\cite{cymulate}, Picus Security~\cite{picus}, SafeBreach~\cite{safebreach}, and SCYTHE~\cite{scythe}); continuous attack-path and exposure management (XM Cyber~\cite{xmcyber} and Pentera~\cite{pentera}); and high-fidelity cyber ranges (SimSpace~\cite{simspace} and Cyberbit~\cite{cyberbit}).
These platforms replay curated adversary playbooks against live or emulated infrastructure to surface control gaps; some now integrate AI assistants for triage and remediation orchestration~\cite{safebreach, pentera}, but the autonomous generation of novel configuration mitigations and their replay-based re-validation remain outside their scope.

\input{Content/tables/environments_table}
\input{Content/tables/ACD_table}

%% file: Content/tables/environments_table.tex
\newcommand{\cmark}{{\color{green!60!black}\checkmark}}
\newcommand{\xmark}{{\color{red!80!black}\texttimes}}
\newcommand{\pmark}{{\color{orange!90!black}\texttildelow}}
\newcommand{\bcmark}{\textbf{\color{green!60!black}\checkmark}}
\newcommand{\bpmark}{\textbf{\color{orange!90!black}\texttildelow}}
\newcommand{\bxmark}{\textbf{\color{red!80!black}\texttimes}}

\begin{table*}[t]
\centering
\footnotesize
\setlength{\tabcolsep}{3pt}
\renewcommand{\arraystretch}{1.15}
\caption{Autonomous cyber defense environments compared on fidelity, attack/defense features, and the validation properties required to evaluate deployable mitigations.
         \cmark~=~supported; \xmark~=~not supported; \pmark~=~partial or unclear.}
\label{tab:environments}
\resizebox{\textwidth}{!}{%
\begin{tabular}{%
  l
  p{1.6cm}                 
  c c c                    
  c c c c                  
  c                        
}
\toprule
& &
\multicolumn{3}{c}{\textbf{Attack / Defence Features}} &
\multicolumn{4}{c}{\textbf{Validation Properties}} & \\
\cmidrule(lr){3-5}\cmidrule(lr){6-9}
\textbf{Platform} &
\textbf{\shortstack{Network\\type}} &
\textbf{\shortstack{Custom.\\topo.}} &
\textbf{\shortstack{Vendor\\dev. imgs}} &
\textbf{\shortstack{Custom.\\attacks}} &
\textbf{Deployable} &
\textbf{\shortstack{Mitigation\\validation}} &
\textbf{\shortstack{Connectivity\\regression}} &
\textbf{\shortstack{Cumulative\\mit.}} &
\textbf{\shortstack{Open\\source}} \\
\midrule
\multicolumn{10}{l}{\textit{Emulation-based}} \\[1pt]
MIRAGE~\citeyearpar{kouremetis2024mirage}
  & Emu., Sim.\textsuperscript{1}
  & \pmark & \xmark & \cmark
  & \xmark & \cmark & \xmark & \xmark
  & \xmark \\

CyGIL\textsuperscript{\dag}~\citeyearpar{li2021cygil}
  & Emulation
  & \pmark & \xmark & \cmark
  & \xmark & \xmark & \xmark & \xmark
  & \xmark \\
\midrule
\multicolumn{10}{l}{\textit{Hybrid (simulation + emulation)}} \\[1pt]
Cyberwheel~\citeyearpar{oesch2024towards}
  & Hybrid
  & \cmark & \pmark & \cmark
  & \xmark & \pmark & \xmark & \xmark
  & \cmark \\
FARLAND~\citeyearpar{molina2021network}
  & Sim., Hybrid
  & \cmark & \xmark & \pmark
  & \pmark & \cmark & \xmark & \xmark
  & \xmark \\
CSLE~\citeyearpar{hammar2022intrusion}
  & Hybrid
  & \cmark & \xmark & \pmark
  & \xmark & \cmark & \xmark & \xmark
  & \cmark \\
NASimEmu\textsuperscript{\dag}~\citeyearpar{janisch2023nasimemu}
  & Hybrid
  & \cmark & \pmark & \pmark
  & \xmark & \xmark & \xmark & \xmark
  & \cmark \\
\midrule
\multicolumn{10}{l}{\textit{Simulation-based}} \\[1pt]
CybORG / CAGE~\citeyearpar{standen2021cyborg,kiely2025exploring,hammar2024optimal}
  & Simulation\textsuperscript{3}
  & \pmark & \xmark & \pmark
  & \xmark & \pmark & \xmark & \xmark
  & \cmark \\
CybORG++~\citeyearpar{emerson2024cyborg++}
  & Simulation
  & \pmark & \xmark & \pmark
  & \xmark & \pmark & \xmark & \xmark
  & \cmark \\
PrimAITE~\citeyearpar{primaite}
  & Simulation
  & \cmark & \xmark & \cmark
  & \xmark & \pmark & \xmark & \xmark
  & \cmark \\
CyberBattleSim~\citeyearpar{msft:cyberbattlesim}
  & Simulation
  & \cmark & \xmark & \pmark
  & \xmark & \pmark & \xmark & \xmark
  & \cmark \\
\midrule
\textbf{COHORT (this work)}
  & \textbf{Emulation}
  & \bcmark & \bcmark & \bcmark
  & \bcmark & \bcmark & \bcmark & \bcmark
  & \bpmark\textsuperscript{2} \\
\bottomrule
\end{tabular}%
}
\smallskip\\
\raggedright\footnotesize
\emph{Deployable} = defenses expressed as real device configuration commands rather than abstract state transitions.\\
\textsuperscript{\dag}Attack-only platform; no defender side.\\
\textsuperscript{1}\url{https://www.mitre.org/our-impact/intellectual-property/cyberlayer}\\
\textsuperscript{2}See Section~\ref{appen:open-science} for artifact and code release status.\\
\textsuperscript{3}CybORG also supports an AWS-based emulation backend; CAGE Challenges, the standard benchmarks built on it, run only the simulation mode.
\end{table*}

%% file: Content/tables/ACD_table.tex
\begin{table*}[t]
\centering
\footnotesize
\setlength{\tabcolsep}{6pt}
\renewcommand{\arraystretch}{1.2}
\caption{Comparison of autonomous cyber defense (ACD) systems and LLM-based network configuration systems.
\cmark~=~supported; \xmark~=~not supported; \pmark~=~partial or limited.}
\label{tab:acd-systems-2}

\begin{tabular}{l l l c c c c}
\toprule

\textbf{System}
& \textbf{Agent}
& \textbf{Environment}
& \textbf{Suggest}
& \textbf{Deploy}
& \textbf{Config-level}
& \textbf{Open source} \\

\midrule

\multicolumn{7}{l}{\textit{LLM-based defensive systems}} \\

AgentNIRS~\citeyearpar{saroui2025agentnirs}
& LLM
& Dataset
& \cmark & \cmark & \cmark
& \cmark \\

CyberAlly~\citeyearpar{kim2025cyberally}
& LLM + KG
& Live cyber range
& \cmark & \xmark & \xmark
& \xmark \\

Castro et al.~\citeyearpar{castro2025large}
& LLM
& Simulation
& \cmark & \pmark & \xmark
& \cmark \\

ShieldGPT~\citeyearpar{wang2024shieldgpt}
& LLM
& Advisory only
& \cmark & \xmark & \xmark
& \cmark \\

\midrule

\multicolumn{7}{l}{\textit{Hybrid RL + LLM systems}} \\

SecurityBot\textsuperscript{\dag}~\citeyearpar{yan2024depending}
& RL + LLM
& Simulation
& \pmark & \pmark & \xmark
& \xmark \\

Loevenich et al.~\citeyearpar{loevenich2025design}
& DRL + LLM
& Simulation
& \pmark & \pmark & \xmark
& \xmark \\

\midrule

\multicolumn{7}{l}{\textit{LLM-based network configuration systems (no adversary in the loop)}} \\

NetConfEval~\citeyearpar{wang2024netconfeval}
& LLM
& Benchmark
& \cmark & \pmark & \cmark
& \cmark \\

GeNet~\citeyearpar{ifland2025genet}
& LLM (multimodal)
& Dataset
& \cmark & \xmark & \cmark
& \cmark \\

LLM-NetCFG~\citeyearpar{lira2024large}
& LLM (local)
& Use-case demo
& \cmark & \cmark & \cmark
& \xmark \\

ConfGen~\citeyearpar{wang2026llm}
& LLM + RAG
& Dataset
& \cmark & \xmark & \cmark
& \xmark \\

\midrule

\multicolumn{7}{l}{\textit{Emulation-based defensive systems}} \\

MIRAGE~\citeyearpar{kouremetis2024mirage}
& Rule-based
& Emulation
& \pmark & \cmark & \xmark
& \xmark \\

\midrule

\textbf{COHORT (this work)}
& \textbf{LLM (multi-agent)}
& \textbf{Emulation}
& \bcmark & \bcmark & \bcmark
& \bpmark\textsuperscript{1} \\

\bottomrule
\end{tabular}

\vspace{2pt}
\raggedright\footnotesize
\emph{Environment} refers to where mitigations are deployed/validated, not training.
\emph{Config-level} = mitigation expressed as device configuration commands rather than abstract state transitions.\\
\textsuperscript{\dag}SecurityBot operates as both attacker and defender; all other rows are defenders.\\
\textsuperscript{1}See Section~\ref{appen:open-science} for artifact and code release status.
\end{table*}

%% file: Content/Threat_Model.tex
\section{Threat Model}
\label{sec:threat-model}
We consider the security of enterprise networks in the post-compromise phase.
The defender's goal is to harden the network so that an observed adversary, when re-executed against the modified topology, makes less progress without breaking legitimate operation.
We evaluate four attack scenarios (Table~\ref{tab:adv_all}): Thief (data theft), Ransomware, DNS Exfiltration, and Lateral Movement.

\paragraph{Threat profile}
COHORT targets \emph{scripted, intelligence-equipped adversaries} that have established a foothold and acquired targeted reconnaissance, replayable from forensic analysis, threat-intelligence feeds, or red-team exercises.
We use Caldera as the replay platform; equivalent BAS or adversary-emulation platforms emitting a comparable operational report can substitute.
Adaptive adversaries that re-plan in response to observed defenses are excluded (Section~\ref{sec:scope-generalization}); fixing the adversary across baseline and post-mitigation runs is what makes the comparison attributable to the deployed mitigation rather than to attacker variation.

\subsection{Adversary Model}
\label{sec:adversary-model}
\paragraph{Knowledge}
The adversary has a foothold on an endpoint and is seeded with targeted intelligence: valid credentials for one LAN host, the IP of an exfiltration target, and sensitive file paths and extensions.
All other facts (additional hosts, services, credentials) must be acquired through on-host reconnaissance and lateral movement during the attack.
The adversary has no visibility into the mitigation framework.

\paragraph{Capabilities}
The adversary operates with normal user privileges and executes MITRE ATT\&CK techniques using tools staged on the foothold host.
The operator control channel is delivered out of band from the production data path and lies outside the defender's mitigation surface, so agent connectivity is never a defensive target; by contrast, exfiltration traffic traverses the production path and is in scope.
Mitigations are credited for blocking exfiltration on the production path, not for severing the control channel, thereby preventing trivial wins and forcing engagement with the actual data flow.
Implementation specifics (Caldera Sandcat agent, GNS3 interface naming) are deferred to \nameref{appen:caldera_linux}.

\subsection{Defender Model}
The defender corresponds to a SOC operator with administrative authority over their own infrastructure.
The framework consumes a structured operational report from executing a known adversary profile (threat-intelligence feeds, the Caldera Stockpile, or red-team exercises) against an emulation of the production network; forensic-grade reconstruction of a real-world breach is not assumed.

\paragraph{Emulation prerequisite}
A high-fidelity emulation of the production network is presupposed as input; automated production-to-emulation generation is an open problem (Section~\ref{dis:open-questions}). The prerequisite is met by organizations that already maintain a staging environment or pre-production lab.

\paragraph{Failure modes}
Perfect execution is not assumed. The agent pipeline can produce invalid configurations, disrupt legitimate connectivity, or generate operationally ineffective controls. The connectivity-preservation gate (Success Criterion, below) excludes mitigations that break connectivity. Connectivity-preserving but ineffective mitigations (rules at the wrong hop, evaluated in the wrong order, or targeting primitives the adversary does not use) surface as ME~$\le$~0 in the replay; Section~\ref{sec:mitigation-categories} characterizes the recurring patterns.

\subsection{Success Criterion}

A mitigation \emph{succeeds} on a run if both of the following hold:
\begin{enumerate}
    \item it strictly reduces the Attack Step Success Rate (ASSR) compared to the unmitigated baseline ($\text{ME} > 0$); and
    \item legitimate connectivity is preserved, verified by two independent probes from the compromised host (a LAN ping to an internal target and an HTTP request to an internet endpoint), both of which must succeed.
\end{enumerate}
Equivalently, letting $\Phi : \mathcal{N} \to \{0,1\}$ denote the connectivity gate, $\Phi(\mathcal{N}) = \mathbf{1}[\text{LAN ping passes} \,\wedge\, \text{HTTP probe passes}]$, the success indicator is $\mathbf{1}[\text{ME}(m;\mathcal{N}_t,\mathcal{A}) > 0 \,\wedge\, \Phi(\mathcal{N}_{t+1}) = 1]$, with $\mathcal{N}_t$, $\mathcal{N}_{t+1}$, and $\mathcal{A}$ as introduced in Section~\ref{sec:framework-formalization}.
We report this binary outcome as the Mitigation Success Rate (MSR), the continuous reduction as the Mitigation Effectiveness (ME), and the compounding effect across stacked mitigations as the Cumulative ME (Section~\ref{sec:eval-metrics}).
The criterion is step-level by design. A step-level signal exposes \emph{which} parts of the kill chain a mitigation disrupts, the granularity needed to drive the iterative critic loop; objective denial follows when the disrupted step lies on the only viable path to the objective (for example, blocking the final T1041 exfiltration step prevents data from leaving the network). Step-level disruption that contributes little to the aggregate effect remains visible through ME and Cumulative ME.

\subsection{Scope and Generalization}
\label{sec:scope-generalization}
The reported numbers generalize to scripted adversaries with observable TTPs whose execution profile can be replayed deterministically. They do not generalize to:
\begin{itemize}
    \item \textbf{Adaptive adversaries} that observe deployed defenses and re-plan. Replay-based validation is non-adaptive by construction (Section~\ref{dis:fixed-adversary}); evaluation against adaptive attackers requires a co-evolving measurement contract, left open (Section~\ref{dis:open-questions}).
    \item \textbf{TTPs absent from MITRE ATT\&CK} or whose execution Caldera cannot emulate.
    \item \textbf{Heterogeneous endpoint OSes.} The current evaluation is Linux-only (Section~\ref{sec:experimental-setup}); whether MSR transfers to Windows or ICS environments is open.
    \item \textbf{Network-device heterogeneity.} The evaluation uses FortiGate, Cisco IOS, and Open vSwitch images (\nameref{appen:firewall}, \nameref{appen:router}, \nameref{appen:switch}); transfer to other vendor families (e.g., Palo Alto, Juniper, Aruba) is open.
    \item \textbf{Operational regression beyond connectivity probes}, including latency, application-layer disruption, and user-visible behavior beyond LAN/internet reachability.
\end{itemize}

\subsection{Out of Scope}

\begin{itemize}
    \item \textbf{Attacks on the framework} or its supporting infrastructure.
    \item \textbf{Initial-access and pre-compromise phases} of the attack lifecycle.
    \item \textbf{Insider threats and physical attacks.}
    \item \textbf{Virtualization infrastructure failures.}
\end{itemize}

%% file: Content/The_Automatic_Mitigation_Framework.tex
\section{COHORT}\label{sec:framework}

\subsection{Framework Scope and Purpose}
COHORT focuses on post-compromise enterprise defense: it proposes network-level mitigations and validates them by replaying the original adversary in an emulated environment, with no production exposure during evaluation.

Fidelity derives from three design choices: (i)~control and data planes run real vendor firmware on the \hyperref[appen:firewall]{firewall}, \hyperref[appen:switch]{switch}, and \hyperref[appen:router]{router}; (ii)~endpoints run as Ubuntu containers; (iii)~Caldera executes real commands, producing genuine traffic and host-level artifacts. We do not model enterprise-scale traffic load, configuration drift, Windows endpoints (Section~\ref{dis:linux-only}), or live synchronization with production; the framework operates as a staging environment on static baseline snapshots. The fidelity claim is scoped to deployable mitigations (Section~\ref{sec:scope-generalization}).

\subsection{Problem Formalization}\label{sec:framework-formalization}

\paragraph{Setting}
Let $\mathcal{N}$ denote the space of network configuration states (device configurations, ACLs, routing policies, host-level configuration). Let $\mathcal{A}$ denote a fixed adversary scenario executed by Caldera, and $\rho$ the procedure that runs $\mathcal{A}$ against a state and returns an operational report $\mathcal{R}$ (ASSR is a function of $\mathcal{R}$, defined in Section~\ref{sec:eval-metrics}). Let $\mathcal{C}$ be the device configuration command space; a \emph{mitigation} is a finite sequence $m \in \mathcal{C}^{*}$, applied via a state transition $\tau : \mathcal{N} \times \mathcal{C}^{*} \to \mathcal{N}$. Let $B \in \mathbb{N}$ be the per-refinement command budget and $K \in \mathbb{N}$ the implementer/critic refinement cap (Section~\ref{sec:experimental-setup}).

\paragraph{Mitigation Cycle}
A \emph{mitigation cycle} starts from a baseline state $\mathcal{N}_t$ and produces a candidate $m$ in three stages: (1)~\emph{replay}: $\mathcal{R}_t = \rho(\mathcal{N}_t, \mathcal{A})$; (2)~\emph{suggest}: the suggester reads $\mathcal{R}_t$ and (per Section~\ref{sec:methodology-agents}, Context Management) its prior suggestions, and proposes a candidate strategy that the implementer expands into commands; (3)~\emph{refine}: the implementer/critic alternation produces a sequence $m^{(0)}, m^{(1)}, \dots, m^{(j)}$ with $j \leq K$ and $|m^{(i)}| \leq B$ for every $i$, halting on critic approval or at $j = K$, with $m:= m^{(j)}$. The post-mitigation state is $\mathcal{N}_{t+1} = \tau(\mathcal{N}_t, m)$. The cycle's binary outcome is the Success Criterion of Section~\ref{sec:threat-model} evaluated on $(\mathcal{N}_t, \mathcal{N}_{t+1}, \mathcal{A})$.

\paragraph{Constraints}
(i)~Per refinement iteration, $|m^{(i)}| \leq B$; the cycle therefore issues at most $K \cdot B$ commands across all iterations (Section~\ref{sec:experimental-setup}).
(ii)~$\mathcal{A}$ is held fixed across the baseline replay $\rho(\mathcal{N}_t, \mathcal{A})$ and the post-mitigation replay $\rho(\mathcal{N}_{t+1}, \mathcal{A})$ used to compute ME (Section~\ref{sec:threat-model}).
(iii)~The suggester is prompted to exclude indicator-based blocking, monitoring-only responses, and allowlists permitting the observed exfiltration protocol; this is a design constraint on the policy class, enforced by prompting rather than by syntactic projection.
(iv)~The single-mitigation environment is reset to a baseline snapshot between cycles; the cumulative environment is not (Section~\ref{sec:cumulative}).

\paragraph{Cumulative Variant}
The cumulative environment maintains a persistent state $\mathcal{N}^{(0)}, \mathcal{N}^{(1)}, \dots$. After the $k$-th candidate $m_k$ is implemented and replayed on $\tau(\mathcal{N}^{(k-1)}, m_k)$, it is accepted, $\mathcal{N}^{(k)} = \tau(\mathcal{N}^{(k-1)}, m_k)$, only if (a) Cumulative ME does not decrease relative to $\mathcal{N}^{(k-1)}$, and (b) $\Phi(\tau(\mathcal{N}^{(k-1)}, m_k)) = 1$; otherwise $\mathcal{N}^{(k)} = \mathcal{N}^{(k-1)}$.

\subsection{Agent Roles and Responsibilities}\label{sec:methodology-agents}
\begin{figure*}[pos=t]
  \centering
  \includegraphics[width=0.95\linewidth]{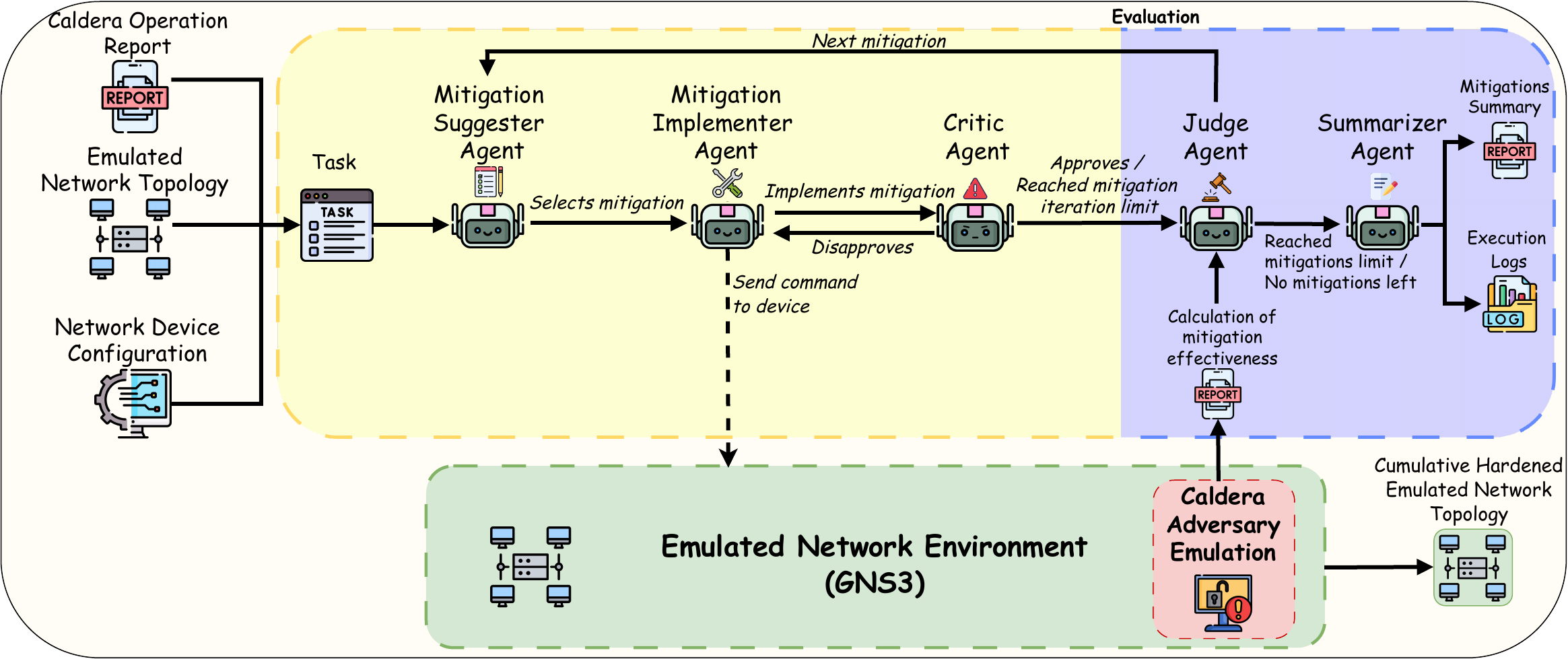}
  \caption{Multi-agent automatic mitigation framework showing the overall architecture and agent interactions.}
    \Description{A flowchart diagram showing the multi-agent automatic mitigation framework. The diagram illustrates the flow from the operational report and network topology through the Task component to the multi-agent system (Mitigation Suggester, Mitigation Implementer, and Critic Agent), which interact with the emulated network environment containing Caldera adversary emulation. The Judge Agent and Summarizer Agent evaluate mitigation outcomes and produce a summary, with feedback loops shown between components.}

\label{fig:system-model}
\end{figure*}

The mitigation workflow is implemented using a team of role-specialized LLM-based agents (Figure~\ref{fig:system-model})~\cite{liu2025foundationagents}.

\paragraph{Mitigation Suggester}
The mitigation suggester analyzes the adversary's operational report, along with the network topology and configuration state, to propose configuration changes that disrupt the observed attack capabilities.
Suggested mitigations must be reusable, generalizable, and expressed at the level of enterprise controls (e.g., segmentation policies, routing constraints, firewall rules).

\paragraph{Mitigation Implementer}
The mitigation implementer translates a selected mitigation into concrete configuration changes within the emulated environment, modifying device configurations or logical topology elements.

\paragraph{Critic}
The critic reviews implemented mitigations prior to offensive replay, identifying configuration errors, policy violations, or deviations from the intended mitigation objective.
When issues are detected, it provides feedback to the mitigation implementer, who revises the configuration accordingly.
This review--revision loop halts on critic approval or at the iteration cap $K$; the empirical lift over the no-critic ablation (Section~\ref{sec:defense-success}) justifies running it.

\paragraph{Judge}
Whereas the critic reviews the implemented configuration prior to attack execution, the judge evaluates the resulting attack outcome.
It replays the attack scenario in the post-mitigation environment using Caldera and compares the resulting attack step success rate (ASSR) to the pre-mitigation baseline; a lower rate indicates a successful mitigation.
The judge also categorizes each mitigation into strategic classes (e.g., \textit{Host Hardening}, \textit{Network-Based Traffic Filtering}).

\paragraph{Summarizer}
After all mitigation cycles are complete, the summarizer reads the judge's per-mitigation evaluations (rather than the full multi-round conversation) and produces a compact retrospective covering the attack scenario, attempted mitigations and outcomes, cross-mitigation observations, connectivity validation results, and the device commands executed by the implementer.

\subsection{Context Management}
Each agent's context scope is tailored to its role. The mitigation suggester receives the history of its own prior suggestions, allowing it to avoid proposing strategies it has already tried; outcomes from earlier mitigations are deliberately excluded both to prevent the suggester from exploiting variations of already-successful strategies and to keep its context (and therefore token cost) bounded across long sessions.
In contrast, the implementer and critic agents receive only the current mitigation context, with no knowledge of prior mitigations, since each mitigation is evaluated independently.

\subsection{Mitigation Cycle}

Each mitigation is evaluated through the following phases:

\begin{description}
  \item[Initialization.] The single-mitigation environment is reverted to the baseline snapshot.
  \item[Attack Execution.] The attack scenario is executed, producing the operational report.
  \item[Mitigation Suggesting.] The suggester proposes a candidate mitigation.
  \item[Mitigation Implementation.] The implementer translates the candidate into device commands.
  \item[Critic Review.] The critic reviews the implementation; refinement runs up to $K$ iterations.
  \item[Evaluation.] The judge replays the attack on the mitigated environment to determine whether the mitigation succeeded.
  \item[Connectivity Validation.] Two Caldera probes (a LAN ping and an internet HTTP check) verify reachability; both must pass for the mitigation to count toward \ac{MSR} (Section~\ref{sec:eval-metrics}).
  \item[Cumulative Replay.] Successful mitigations are replayed onto the persistent cumulative project (Section~\ref{sec:cumulative}).
\end{description}

\subsection{Single-Agent Baseline}\label{sec:single-agent}

\begin{figure}[pos=htbp]
  \centering
  \includegraphics[width=0.95\linewidth]{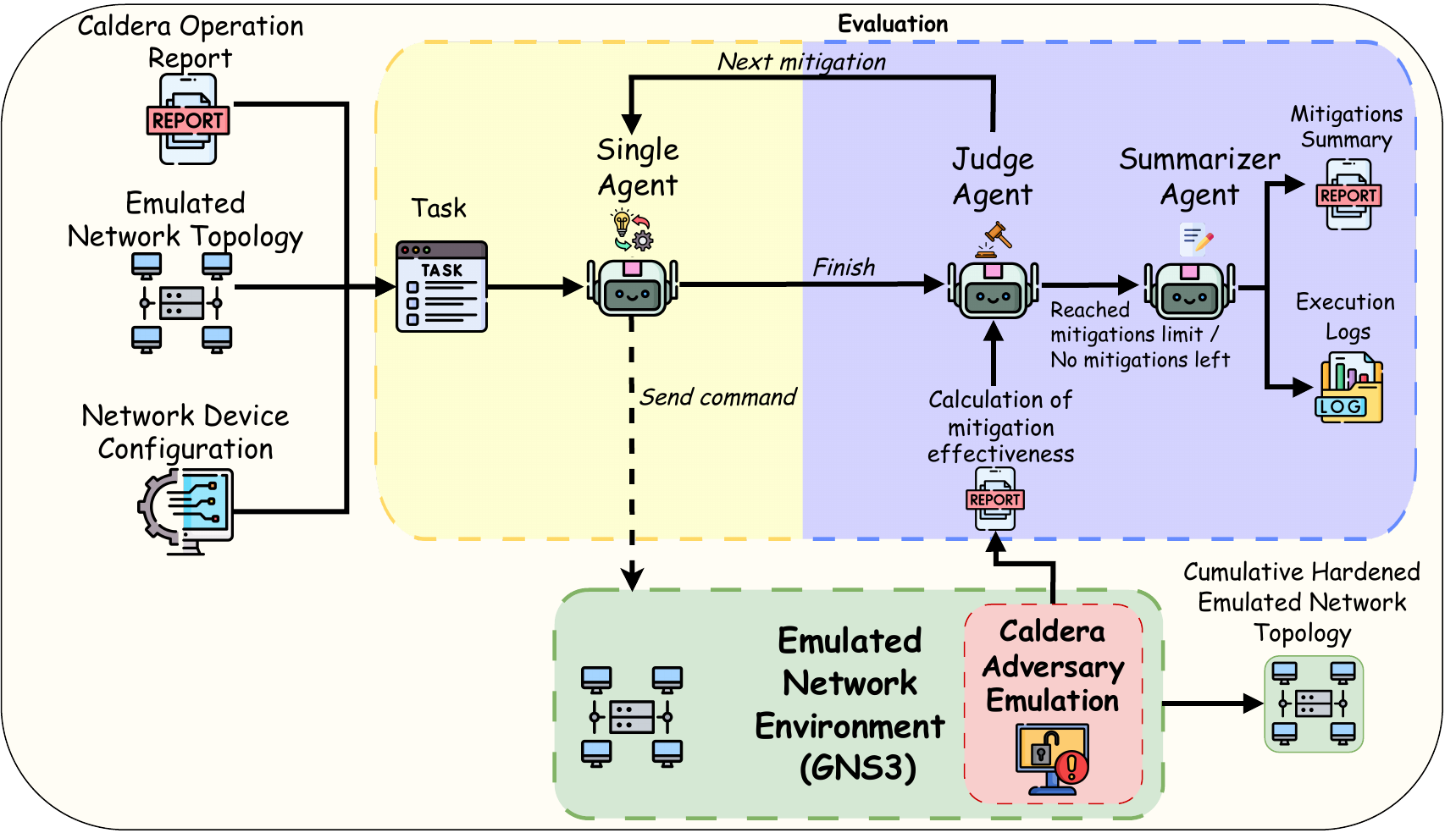}
  \caption{Single-agent baseline workflow for mitigation suggestion, implementation, and self-validation.}
  \Description{A simplified flowchart showing the single-agent mitigation workflow. A single LLM agent receives the operational report and network topology, proposes and implements a mitigation by sending commands to devices, and is evaluated by the Judge Agent and Summarizer Agent.}
\label{fig:single-agent}
\end{figure}

To isolate the contribution of the multi-agent architecture, we evaluate a single-agent baseline that consolidates the Suggester, Implementer, and Critic roles into a single LLM; the Judge and Summarizer roles are unchanged.
The single agent shares the same model (\hyperref[appen:llm]{GPT-5.4~mini}), inputs (operational report, network topology, configuration state), and device-configuration access as the multi-agent system.
Figure~\ref{fig:single-agent} shows the simplified workflow.

\subsection{Cumulative Mitigation Evaluation}\label{sec:cumulative}

\begin{figure*}[pos=t]
  \centering
  \includegraphics[width=0.95\linewidth]{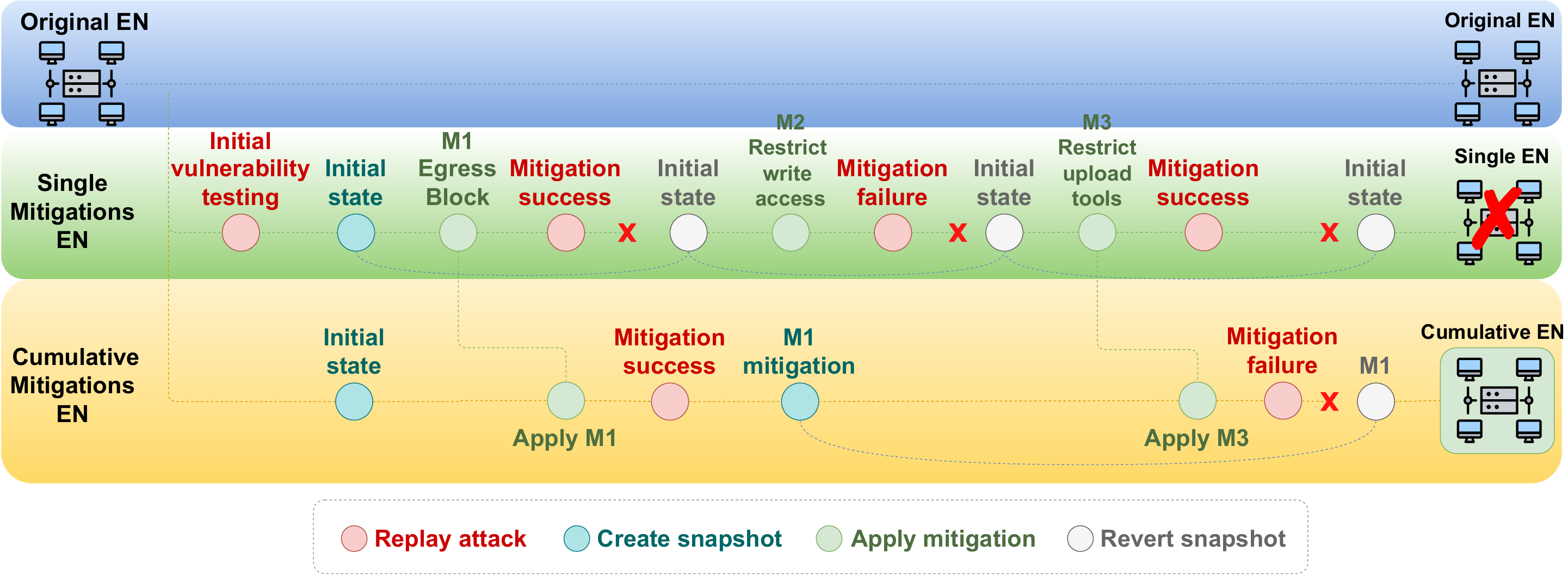}
  \caption{Evaluation workflow showing the parallel per-mitigation evaluations (independent, rolled-back) and the cumulative evaluation (persistent cumulative project with sequential mitigation replay). A demo video walking through this diagram can be found in Appendix~\ref{appen:open-science}.}
  \Description{A workflow diagram showing two parallel evaluation paths. The top path shows individual mitigations tested on a clean baseline with rollback. The bottom path shows the cumulative project where approved mitigations accumulate sequentially. A legend shows success, validation failure, and rejection states.}
\label{fig:gitflow}
\end{figure*}

Per-mitigation evaluation reveals individual effectiveness but not how defenses interact when stacked. 
The cumulative evaluation (Figure~\ref{fig:gitflow}) captures these compound effects: a mitigation that passed the per-mitigation Evaluation with connectivity preserved is replayed onto a persistent \emph{cumulative project}, an emulated network instance that accumulates approved defenses.
After each cumulative replay, the mitigation is accepted into the project only if it (i) does not regress the Cumulative Mitigation Effectiveness (Cumulative ME, Section~\ref{sec:eval-metrics}) and (ii) preserves basic connectivity; otherwise the project is rolled back to the previously accepted state.
Because acceptance requires non-regression rather than strict improvement, non-interfering mitigations are retained even when they add no measurable ASSR reduction, yielding a \emph{defense-in-depth} posture.

\paragraph{Failure modes}
Each role has a characteristic failure shape. The suggester proposes generalizable controls that do not bind the adversary to the actual primitive, for example, disabling a specific binary while equivalent alternatives remain available. The implementer produces syntactically valid configurations whose effective scope is wrong: rules placed at the wrong hop, evaluated in the wrong order, or scoped to a path that does not intersect the attacker's read or write targets. The critic catches errors visible in the configuration text (syntax errors, policy violations, deviations from stated intent) but cannot distinguish a syntactically correct rule that fails in placement from one that succeeds, since placement effects only surface during replay. Section~\ref{sec:mitigation-categories} reports the empirical breakdown by mitigation category.

%% file: Content/Evaluation.tex
\section{Evaluation}\label{sec:evaluation}

\subsection{Experimental Setup}\label{sec:experimental-setup}

All agent roles (mitigation suggester, mitigation implementer, and critic) use \hyperref[appen:llm]{GPT-5.4~mini} (via Azure OpenAI) with the default sampling temperature of 1.0. The Single-Agent baseline (Section~\ref{sec:single-agent}) uses a single GPT-5.4~mini instance to play all roles in a single conversation. Caldera replay is deterministic given a fixed scenario and network state, so the five repetitions per condition (below) capture only the variance attributable to LLM sampling at this temperature.

\paragraph{Network Topologies}\label{sec:network-topology}
Three representative enterprise network topologies of increasing complexity are used (Figures~\ref{fig:small_enterprise}--\ref{fig:large_enterprise}). The \hyperref[fig:small_enterprise]{\textbf{small enterprise}} topology consists of one FortiGate firewall and a single layer-2 switch connecting workstations and servers, representing a low-complexity baseline environment. The \hyperref[fig:medium_enterprise]{\textbf{medium enterprise}} topology introduces a two-segment network behind a centralized firewall, with two distribution routers and two layer-2 switches enabling inter-segment routing and realistic VLAN-based segmentation. The \hyperref[fig:large_enterprise]{\textbf{large enterprise}} topology adopts a hierarchical three-tier architecture: five core routers interconnect four floor segments, each terminated by a dedicated floor router, a floor-level firewall, and a floor L3 switch, with twelve departmental access switches distributed across the floors. These three topologies approximate small-office, mid-market-branch, and multi-department enterprise-campus deployments, respectively.

\paragraph{Attack Scenarios}\label{sec:attack-scenarios}
We focus on adversary objectives that pose concrete, recurring risks to enterprise environments. All four attack scenarios are implemented as Caldera adversaries with MITRE ATT\&CK-aligned abilities. Together, these adversaries span distinct tactical objectives across the attack lifecycle:

\begin{itemize}
    \item \textbf{Thief}, a multi-stage attack involving local file discovery, staging, archiving, and exfiltration over the C2 channel, stressing access controls and egress constraints (adapted from the \textit{Thief} adversary pack in the \href{https://github.com/mitre/stockpile/blob/master/data/adversaries/packs/1a98b8e6-18ce-4617-8cc5-e65a1a9d490e.yml}{MITRE Caldera Stockpile plugin}~\cite{Stockpile,caldera-thief-blog}).
    \item \textbf{Ransomware}, which performs system reconnaissance and file discovery on the compromised host, then pivots via SSH to a dedicated victim container~(\nameref{appen:ssh_victim}) to encrypt sensitive files with AES, deploy a ransom note, and establish cron-based persistence, before exfiltrating the encryption key to C2.
    \item \textbf{DNS Exfiltration}, which base64-encodes sensitive file contents and exfiltrates them as DNS subdomain queries to bypass HTTP egress controls, then removes artifacts.
    \item \textbf{Lateral Movement}, which scans the LAN for SSH services, brute-forces credentials against a dedicated victim container~(\nameref{appen:ssh_victim}), pivots in via SSH, enumerates the remote host and collects files, then exfiltrates them to C2.
\end{itemize}
All attacks include a dedicated C2 communication link (see Section~\ref{sec:adversary-model}) and end with the ability to validate LAN and internet connectivity. The C2 server orchestrates attacks against a compromised \nameref{appen:caldera_linux} running the Caldera Sandcat agent; the Lateral\_Movement and Ransomware adversaries additionally target a dedicated \nameref{appen:ssh_victim} that serves as the SSH lateral movement destination. Per-step breakdowns of each adversary, including tactic and MITRE ATT\&CK technique identifiers, are provided in the \nameref{appen:adversary-steps} appendix (Table~\ref{tab:adv_all}).

\paragraph{Experimental Conditions}\label{sec:experimental-conditions}
To disentangle the contribution of role specialization from that of iterative critique, we evaluate three runtime conditions that share the same model (\hyperref[appen:llm]{GPT-5.4~mini}), the same Caldera adversaries, and the same empirical validation pipeline:

\begin{itemize}
    \item \textbf{Single-Agent.} A single GPT-5.4~mini instance plays the mitigation suggester, implementer, and critic roles within one conversation (Section~\ref{sec:single-agent}). There is no iterative critique loop: the agent proposes, implements, and self-checks in a single linear trajectory.
    \item \textbf{Multi-Agent without Critic.} The multi-agent graph is executed with role-specialized suggester and implementer agents, and the implementation--critique iteration is constrained to a single pass, so the critic's feedback is never fed back into a revision. The implementer's initial output is taken as-is, without acting on the critic's advice.
    \item \textbf{Multi-Agent.} The full framework, including the critic and up to five implementation--critique iterations per mitigation.
\end{itemize}

Each of the 12 scenarios (4 attacks $\times$ 3 topologies) is repeated five times per condition, with a limit of 10 mitigations per run, up to 5 implementation--critique iterations per mitigation, and a budget of 20 device commands per implementation cycle, instantiating $K=5$ and $B=20$ in the formalization of Section~\ref{sec:framework-formalization}. Across the three conditions we collected 1\,782 mitigation attempts across 179 runs: 600 mitigations (60 runs) Single-Agent, 599 (60 runs) Multi-Agent without Critic, and 583 (59 runs) Multi-Agent. Sub-cap mitigation counts reflect the multi-agent graph's early-exit edge when the suggester proposes no further distinct candidates; one Multi-Agent replicate was excluded after its summary file failed to write.

In all three conditions, the suggester operates under an additional evaluation-time restriction: candidate mitigations may not alter the network topology (no adding, removing, or substituting devices, links, or services). The restriction is motivated by deployability and by frequent agent failures observed when attempting topology-altering changes during early system testing.

\subsection{Evaluation Criteria}\label{sec:eval-metrics}

Our evaluation is grounded in operational outcomes: every mitigation is validated by replaying the attack on the mitigated network and comparing the resulting ASSR to the pre-mitigation baseline.

\paragraph{Mitigation Effectiveness (ME)}
Mitigation effectiveness quantifies the reduction in the Attack Step Success Rate (ASSR) achieved by a deployed mitigation, with the symbols of Section~\ref{sec:framework-formalization}:
\begin{equation}
\text{ME}(m;\mathcal{N}_t,\mathcal{A}) = \text{ASSR}(\mathcal{N}_t,\mathcal{A}) - \text{ASSR}(\tau(\mathcal{N}_t,m),\mathcal{A})
\end{equation}
where $\text{ASSR}(\mathcal{N},\mathcal{A})$ is the ratio of successfully executed Caldera abilities to attempted abilities when adversary $\mathcal{A}$ is replayed against state $\mathcal{N}$, with each ability scored as success, failure, or timeout. ME~$=$~$\text{ASSR}(\mathcal{N}_t,\mathcal{A})$ signifies complete mitigation (post-mitigation ASSR reduced to zero) and ME~$=$~0 indicates no effect. ME can also be negative, where the attempt actively worsens defense, and the attack succeeds at a higher rate than at baseline.

\paragraph{Mitigation Success Rate (MSR)}
MSR is the probability that a proposed mitigation works: the fraction of attempts that both disrupt the attack ($\text{ME}>0$) and preserve legitimate connectivity ($\Phi=1$, Section~\ref{sec:threat-model}).
\begin{equation}
\text{MSR} = \frac{N_{\text{ME}>0 \,\wedge\, \Phi=1}}{N_{\text{total}}} \times 100\%
\end{equation}
where $N_{\text{total}}$ is the total number of mitigation attempts and the numerator counts attempts whose $i$-th cycle satisfies $\text{ME}(m_i;\mathcal{N}_{t_i},\mathcal{A}) > 0$ and $\Phi(\tau(\mathcal{N}_{t_i},m_i)) = 1$.

\paragraph{Cumulative Mitigation Effectiveness ($\text{ME}_{\text{cumulative}}$)}
While ME and MSR measure each mitigation independently (with rollback between attempts), the cumulative evaluation measures how defenses compound when deployed together. Cumulative ME after $k$ approved mitigations is
\begin{equation}
\text{ME}_{\text{cumulative},k} = \text{ASSR}(\mathcal{N}^{(0)},\mathcal{A}) - \text{ASSR}(\mathcal{N}^{(k)},\mathcal{A})
\end{equation}
where $\mathcal{N}^{(0)},\dots,\mathcal{N}^{(k)}$ is the sequence of cumulative-project states defined by the acceptance rule of Section~\ref{sec:framework-formalization} (Cumulative Variant).

Cumulative ME accumulates as a step function: each accepted mitigation either reduces ASSR by a discrete amount or has no incremental effect on top of the defenses already deployed on the cumulative project. Aggregate cumulative-ME behavior across all runs is reported in Figures~\ref{fig:cum_mean_cond} (by condition) and~\ref{fig:cum_topo} (by topology).

\subsection{Results}\label{sec:results}

\paragraph{MSR Results}\label{sec:defense-success}

\begin{figure}[pos=tb]
    \centering
    \includegraphics[width=\linewidth]{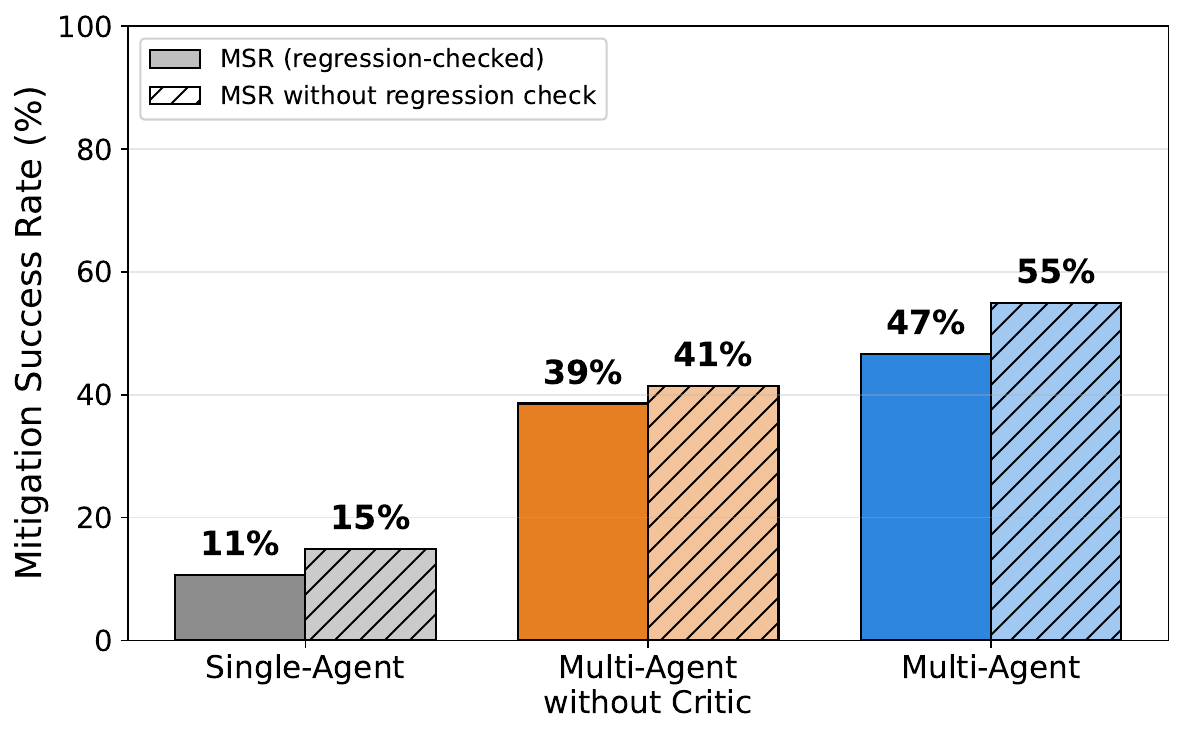}
    \caption{MSR by runtime condition, pooled across attacks and topologies. The hatched bar shows the rate without the connectivity-regression check.}
    \label{fig:msr_by_condition}
\end{figure}

Figure~\ref{fig:msr_by_condition} reports the headline MSR (per-attack split in Appendix~\ref{appen:msr-by-attack}). 
Across all attacks, MSR rises from \textbf{\msrSingle\%} (Single-Agent) to \textbf{\msrMultiNoCritic\%} (Multi-Agent without Critic) to \textbf{\msrMultiCritic\%} (Multi-Agent). 
Every pairwise difference is statistically significant under Holm-Bonferroni correction (full pairwise statistics in Appendix~\ref{appen:pairwise-stats}).
Decomposed against the Single-Agent baseline, role specialization alone (Multi-Agent without Critic) accounts for roughly three-quarters of the total gain over Single-Agent, and the iterative critique loop contributes the remaining one-quarter.
The same ranking holds at the run level: across the run's up-to-ten sequential attempts, Multi-Agent produces at least one working mitigation in $\discoveryMultiCriticPlateau\%$ of runs and Multi-Agent without Critic in $\discoveryMultiNoCriticPlateau\%$, against $\discoverySinglePlateau\%$ for Single-Agent.
Connectivity preservation moves non-monotonically: Multi-Agent without Critic improves it slightly over Single-Agent ($\connMultiNoCritic\%$ vs.\ $\connSingle\%$), while the critic trades it down to $\connMultiCritic\%$ in exchange for higher MSR. The critic's more aggressive controls occasionally break a probe while still blocking the attack.

\begin{keyresult}
\textbf{Multi-agent decomposition lifts MSR $\msrMultiCriticOverSingle\times$ over a Single-Agent baseline.}
Single-Agent $\msrSingle\%$, Multi-Agent without Critic $\msrMultiNoCritic\%$, Multi-Agent $\msrMultiCritic\%$.
Role specialization accounts for roughly three-quarters of the lift; the iterative critique loop accounts for the remaining quarter.
\end{keyresult}

\paragraph{Cost}\label{sec:cost}

The lift carries a cost (Table~\ref{tab:cost_stats}, \nameref{appen:cost}). Per mitigation, Multi-Agent averages \$0.47 and 124~s of wall-clock time, compared to \$0.09 and 60~s for Single-Agent ($\approx$5$\times$ cost and 2$\times$ wall-clock time). Multi-Agent without Critic falls in between at \$0.22 and 96~s. Total experiment cost across all 1\,782 mitigations was \$464.

\paragraph{Cumulative Mitigation Effectiveness}\label{sec:cumulative-effectiveness}

\begin{figure}[pos=tb]
    \centering
    \includegraphics[width=\linewidth]{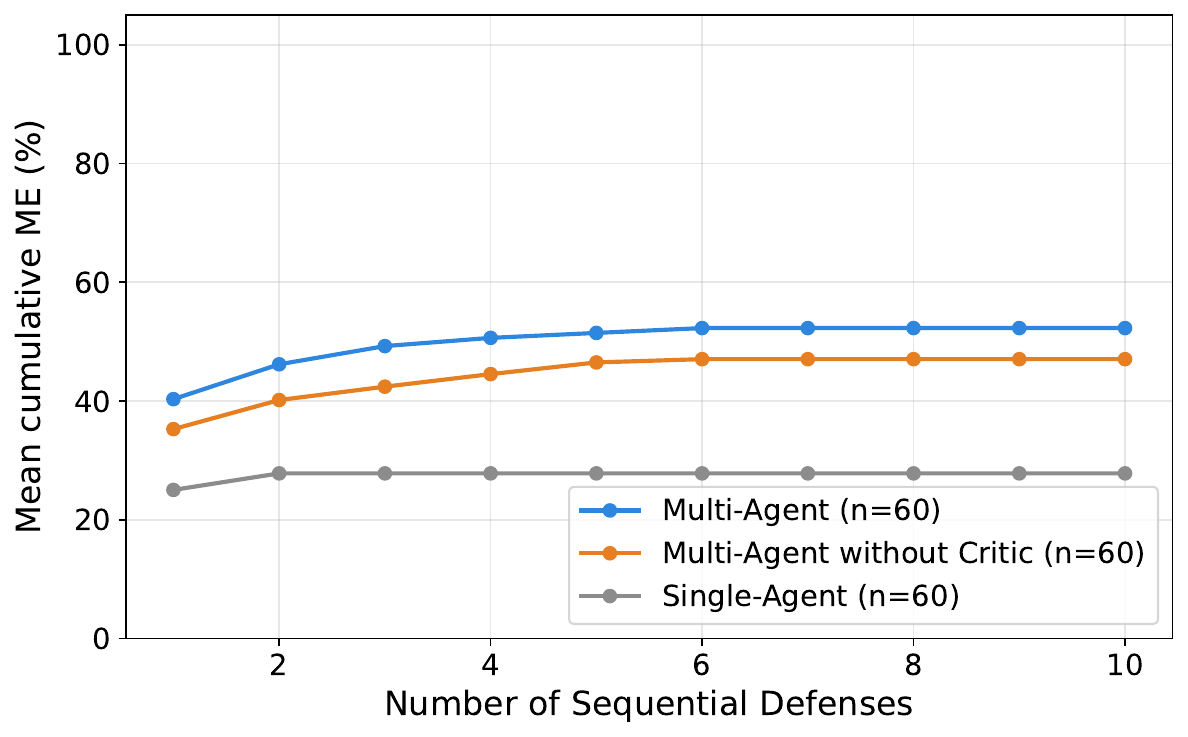}
    \caption{Cumulative ME by condition, pooled across attacks.}
    \label{fig:cum_mean_cond}
\end{figure}

Figure~\ref{fig:cum_mean_cond} characterizes how defenses compound when stacked. Multi-Agent plateaus at $\approx\!\plateauMultiCritic\%$ cumulative ME pooled across attacks, Multi-Agent without Critic at $\approx\!\plateauMultiNoCritic\%$, and Single-Agent at $\approx\!\plateauSingle\%$.

\begin{figure}[pos=tb]
    \centering
    \includegraphics[width=\linewidth]{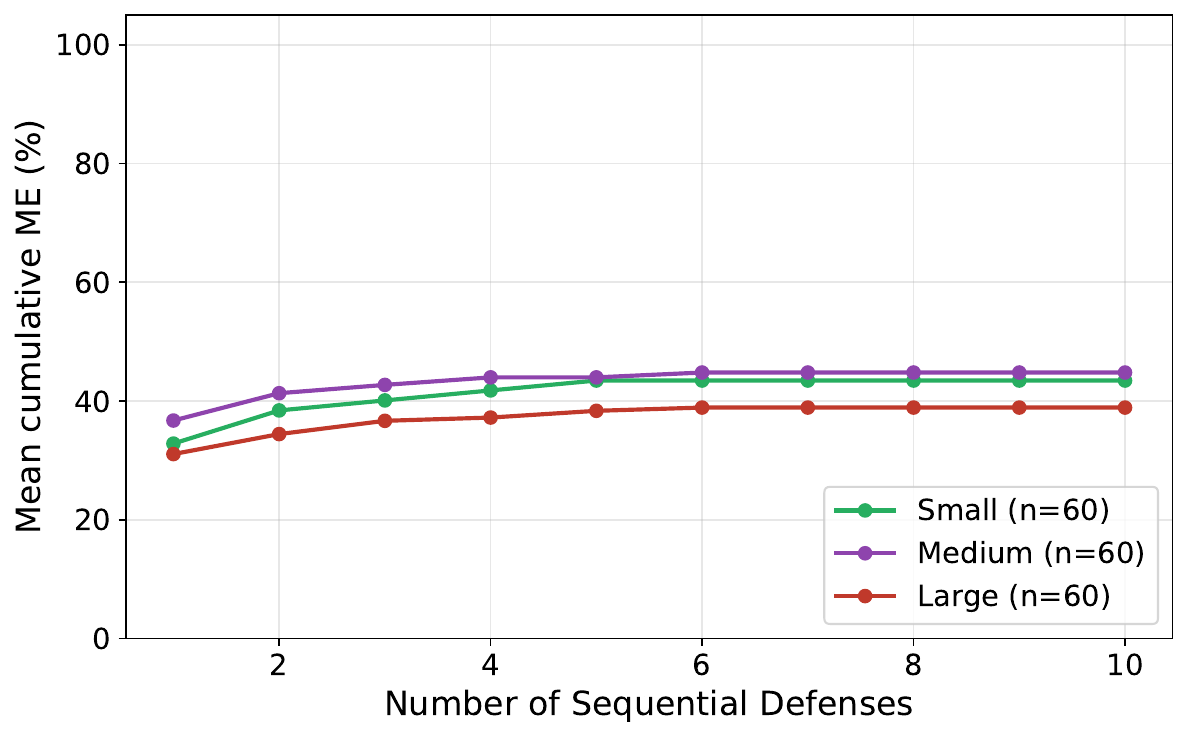}
    \caption{Cumulative ME by topology, all conditions pooled.}
    \label{fig:cum_topo}
\end{figure}

The topology view (Figure~\ref{fig:cum_topo}) shows that, once enough mitigations are stacked, the final cumulative ME is comparable across small, medium, and large enterprise networks. The topology slows the rate of ME accumulation more than it bounds the achievable ceiling. The cumulative track follows a greedy policy: each accepted mitigation is permanently retained on the cumulative project, and earlier choices are not revisited as later mitigations are added.

\paragraph{MSR by Topology}\label{sec:topology-effect}

\begin{figure}[pos=tb]
    \centering
    \includegraphics[width=0.75\linewidth]{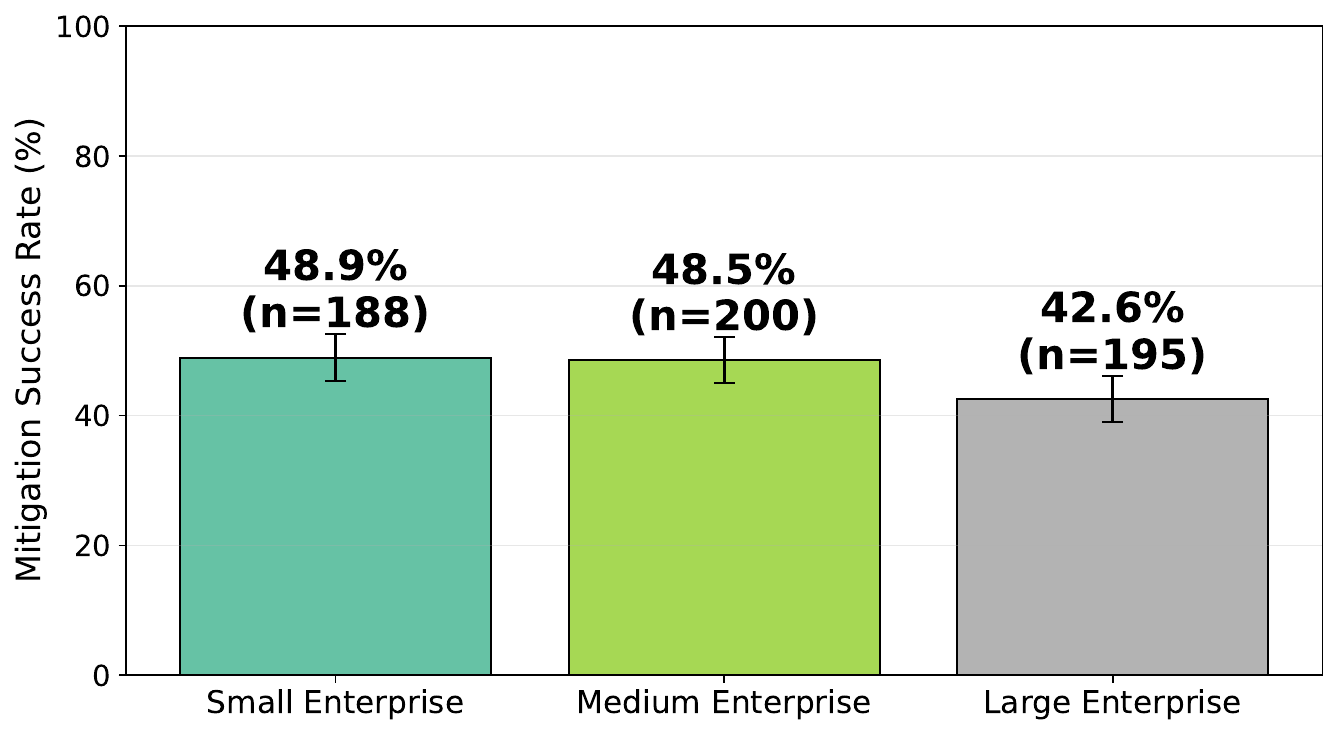}
    \caption{MSR by enterprise topology (small/medium/large) for the Multi-Agent condition. Error bars indicate the standard error of the mean.}
    \label{fig:msr_by_topology}
\end{figure}

Figure~\ref{fig:msr_by_topology} reports MSR by topology for the Multi-Agent condition. We do not detect a topology effect: small enterprise $\msrTopoSmall\%$ ($n=\nTopoSmall$), medium $\msrTopoMedium\%$ ($n=\nTopoMedium$), large $\msrTopoLarge\%$ ($n=\nTopoLarge$); Welch's ANOVA is non-significant ($p=\anovaTopoP$), with all pairwise comparisons non-significant under Holm correction (smallest $p=\anovaTopoPPairMin$). The cumulative-ME view (Figure~\ref{fig:cum_topo}) is consistent: all three topologies plateau by the fourth sequential defense, at $\approx\!\plateauTopoSmall\%$ (small), $\approx\!\plateauTopoMedium\%$ (medium), and $\approx\!\plateauTopoLarge\%$ (large).

\begin{keyresult}
\textbf{No detectable topology effect.} Differences between small ($\msrTopoSmall\%$), medium ($\msrTopoMedium\%$), and large ($\msrTopoLarge\%$) topologies are not statistically significant.
\end{keyresult}

\paragraph{Mitigation Categories}\label{sec:mitigation-categories}

\begin{figure}[pos=tb]
    \centering
    \includegraphics[width=\linewidth]{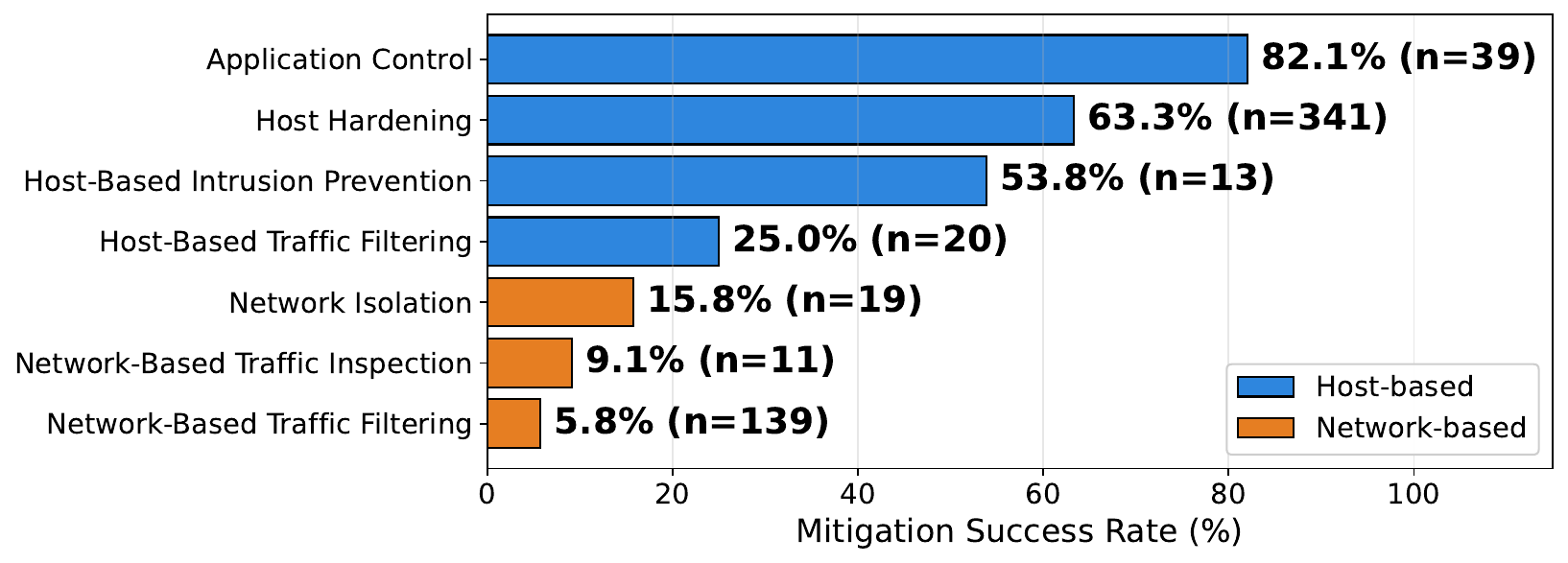}
    \caption{MSR by mitigation category for the Multi-Agent condition, pooled across topologies. Category labels are assigned for each mitigation and normalized to a shared reporting vocabulary (Table~\ref{tab:category_definitions}). Error bars indicate the standard error of the proportion.}
    \label{fig:mitigation_categories}
\end{figure}

To characterize \emph{which} classes of defense the system produces, we group mitigations into high-level categories based on their defensive approach. These categories can be broadly grouped into network-centric controls (enforced at the network perimeter or within the infrastructure), host-centric controls (applied directly to endpoints), and reactive/operational measures. Figure~\ref{fig:mitigation_categories} presents the per-category MSR for the Multi-Agent condition, pooled across topologies; Table~\ref{tab:category_definitions} gives definitions, examples, and counts.

\input{Content/tables/category_definitions.tex}

Two high-level patterns are visible. First, \textbf{host-centric controls dominate in both volume and effectiveness}: Host Hardening (the largest single bucket at $n=\catHostHardeningN$) achieves $\catHostHardeningMSR\%$ MSR, Application Control $\catAppControlMSR\%$ ($n=\catAppControlN$), and Host-Based Intrusion Prevention $\catHBIPSMSR\%$ ($n=\catHBIPSN$, too few attempts to rank reliably against the larger buckets). Second, \textbf{network-based filtering dominates in volume but not in effectiveness}: Network-Based Traffic Filtering is the second-largest bucket ($n=\catNetTrafficFilterN$) but only $\catNetTrafficFilterMSR\%$ of its attempts reduce the attack. Inspection of the failed network-filtering mitigations shows a recurring failure mode: the control is syntactically correct and aligned with best-practice designs (VLAN segmentation, inter-VLAN blocks, router ACLs, egress firewall rules) but is not enforced at the correct point in the network hierarchy, is evaluated in the wrong rule order (e.g., a deny rule after an allow rule on FortiGate), or fails to account for hidden intermediaries on the routed path. The failed host-centric mitigations exhibit an analogous targeting error: the control is correctly applied but does not bind to the attacker's actual primitive. The dominant patterns are disabling a specific binary (e.g., \texttt{find}, \texttt{tar}, \texttt{python3}) while the attacker's tooling has equivalent alternates; tightening filesystem permissions on a path that does not intersect the attack's actual read or write targets; and hardening SSH authentication when the validated attack path is already authenticated.

\begin{keyresult}
\textbf{Host-centric controls dominate; network filtering looks right on paper but rarely fires.} Host Hardening, the largest bucket at $n=\catHostHardeningN$, reaches $\catHostHardeningMSR\%$ MSR, while Network-Based Traffic Filtering, the second-largest by volume at $n=\catNetTrafficFilterN$, reduces the attack only $\catNetTrafficFilterMSR\%$ of the time. Network filtering fails not from wrong intent but from rules placed at the wrong hop or evaluated in the wrong order.
\end{keyresult}

%% file: Content/tables/category_definitions.tex
\begin{table*}[t]
\centering
\caption{Mitigation categories, MITRE D3FEND mappings, and per-category MSR for the Multi-Agent condition, pooled across topologies.}
\label{tab:category_definitions}
\footnotesize
\setlength{\tabcolsep}{4pt}
\begin{tabularx}{\textwidth}{l >{\hsize=0.75\hsize\linewidth=\hsize\raggedright\arraybackslash}X >{\hsize=1.25\hsize\linewidth=\hsize\raggedright\arraybackslash}X l rr}
\toprule
\textbf{Category} & \textbf{Definition} & \textbf{Examples} & \textbf{D3FEND} & \textbf{N} & \textbf{MSR (\%)} \\
\midrule
\makecell[tl]{Application\\Control} & Per-account execute restrictions on specific applications & Per-user execute-denylist of attacker tooling (\texttt{sshpass}, \texttt{nmap}, \texttt{dig}, base64 + Python exfil staging); allowlists permitting only business-required network utilities & \makecell[tl]{D3-EAL\\D3-EDL} & 39 & 82.1 \\
\makecell[tl]{Host\\Hardening} & Host-level reduction of attacker capability via permission and credential restriction & \texttt{chmod -x} on attacker-used utilities (\texttt{sshpass}, \texttt{curl}, \texttt{openssl}, \texttt{dig}); SSH key-only authentication with password auth disabled; per-user write/permission tightening on staging paths & \makecell[tl]{D3-LFP\\D3-SPP\\D3-UAP} & 341 & 63.3 \\
\makecell[tl]{Host-Based\\Intrusion Prevention} & Active host-runtime defenses that detect and respond to in-progress attack actions & \texttt{fail2ban} SSH event-thresholding and account lockout on authentication-failure bursts; cron-based watchdog kills of reconnaissance binaries & \makecell[tl]{D3-ANET\\D3-AL\\D3-PT} & 13 & 53.8 \\
\makecell[tl]{Host-Based\\Traffic Filtering} & Endpoint-local egress firewall enforcement & Host \texttt{iptables}/\texttt{nftables} egress rules denying outbound HTTPS to C2 destinations while preserving LAN ICMP and HTTP validation & \makecell[tl]{D3-OTF\textsuperscript{h}} & 20 & 25.0 \\
\makecell[tl]{Network\\Isolation} & Network-segmentation controls limiting lateral reach & FortiGate / Open vSwitch inter-VLAN ACLs; east-west SSH restrictions and floor-router segmentation between department LANs & \makecell[tl]{D3-BDI\\D3-NI} & 19 & 15.8 \\
\makecell[tl]{Network-Based\\Traffic Inspection} & Application-layer inspection at network devices & FortiGate DNS-security inspection blocking long / high-entropy subdomains and query-burst anomalies & \makecell[tl]{D3-DNSTA\\D3-DNSDL\\D3-PMAD} & 11 & 9.1 \\
\makecell[tl]{Network-Based\\Traffic Filtering} & Perimeter firewall egress and proxy controls & FortiGate outbound HTTPS denies to C2 endpoints; OpenWrt UCI rules blocking DNS-tunnel ports (53, 8853); destination/IP allowlists & \makecell[tl]{D3-OTF\\D3-NTF\\D3-FRDDL} & 139 & 5.8 \\
\bottomrule
\end{tabularx}
\vspace{2pt}
{\footnotesize\raggedright\textsuperscript{h} D3-OTF enforced on the endpoint rather than at a network device.\par}
\end{table*}

%% file: Content/Discussion.tex
\section{Discussion}\label{sec:discussion}

\subsection{Attribution and the Optimization-Evaluation Gap}\label{dis:fixed-adversary}

Caldera replays each adversary scenario unchanged across pre- and post-mitigation runs (Section~\ref{sec:threat-model}).
Holding the attack fixed makes per-mitigation ASSR change attributable to the deployed mitigation rather than to concurrent attacker adaptation.
The same property creates an optimization-evaluation gap: the suggester reads the operational report from the same replay that the judge later scores, so candidates can, in principle, be tuned to the evaluation steps.

Several design properties bound this gap.
The success criterion pairs ME~$>$~0 with a connectivity check (Section~\ref{sec:eval-metrics}), so trivial wins like severing all egress fail outright.
Mitigations are credited against the attack's production-path actions, not the operator's C2 channel (Section~\ref{sec:adversary-model}), which forces engagement with the data flow rather than the attacker's infrastructure.
The suggester is restricted to reusable enterprise-level controls, with indicator-based blocking and monitoring-only responses excluded by prompt (Section~\ref{sec:framework-formalization}).
The framework's outputs are reviewed by engineers before deployment, not auto-applied, so final acceptance rests on human review rather than on the metric.

\subsection{Failure Analysis: Where Mitigations Miss}\label{dis:failure-analysis}

Host Hardening reaches $\catHostHardeningMSR\%$ MSR ($n=\catHostHardeningN$) and Application Control $\catAppControlMSR\%$ ($n=\catAppControlN$), while Network-Based Traffic Filtering reaches only $\catNetTrafficFilterMSR\%$ ($n=\catNetTrafficFilterN$) (Section~\ref{sec:mitigation-categories}).
Host-local controls succeed because the implementer has the resident context (file, process, user, binary) it acts on, so misses concentrate in cases where the chosen primitive does not bind the attacker's: disabling \texttt{find} when \texttt{tar} suffices, hardening SSH when the validated path is already authenticated. 
Network controls fail because rules are reasoned about as syntactic objects rather than bound to a routed path: rules at the wrong hop, allow before deny, or scoped past hidden intermediaries.

The cumulative-ME plateau ($\plateauMultiCritic\%$ pooled, $\plateauTopoSmall\%$, $\plateauTopoMedium\%$, and $\plateauTopoLarge\%$ across topology sizes; ANOVA $p=\anovaTopoP$) is consistent with the same bound: stacked greedy defense saturates against the attack steps that no single host- or network-side control covers under our connectivity contract.

\subsection{Linux-Only Endpoints}\label{dis:linux-only}

GNS3 supports Windows endpoint images, but Windows guests are substantially more compute-intensive than Linux containers in emulation, so we restrict the evaluation to Linux endpoints. 
Whether the reported MSR transfers to Windows-hosted environments is open, and the exposure is uneven across mitigation categories. 
Host Hardening and Application Control depend on platform-specific binaries, policy primitives, and default services that differ substantially between Linux and Windows; Network-Based Traffic Filtering operates on the same TCP/IP path and inherits the same placement-bound constraints (Section~\ref{dis:failure-analysis}).

\subsection{Ability-Level Granularity}\label{dis:granularity}

MSR and ME are reported at the granularity of Caldera abilities. 
When a single ability bundles multiple discrete actions, intra-ability disruption is not resolved: a partially broken ability still counts as a successful step. 
This biases reported disruption downward rather than upward, since a mitigation that thwarts most but not all of a bundled ability fails the ME~$>$~0 check. 

\subsection{Open Questions}\label{dis:open-questions}

First, the cumulative project produces a persistent hardened network state at the end of each attack's track.
Reusing it as the starting topology for a different adversary would let the framework measure cross-attack transfer directly: whether mitigations stacked against one attack continue to disrupt a second, distinct adversary. Iterating this handoff across multiple adversaries extends the cumulative track into a sequential multi-attack evaluation, in which each attack inherits the hardened state left by the previous one.

Second, this cumulative track currently uses a greedy stacking policy in which each accepted mitigation is retained on the cumulative project and never revisited. 
Non-greedy variants (lookahead over candidate mitigations, beam search across cumulative-project trajectories, or a revisit-and-replace operator that swaps earlier choices when later mitigations subsume them) would allow the framework to probe whether the observed cumulative-ME plateau is local to the greedy policy or near a global ceiling for stacked defense.

Third, grounding the suggester in MITRE D3FEND or organization-specific incident-response playbooks would constrain proposals to a vetted defensive vocabulary, align autonomous mitigations with approved response actions, and produce mitigations whose primitives are immediately recognizable to security operations staff.

Fourth, automated generation of emulated topologies from production network configurations would reduce the manual configuration burden associated with high-fidelity replicas. 
While LLMs can assist with such setup~\cite{ifland2025genet, wang2024netconfeval}, fully automated production-to-emulation generation remains an open engineering challenge, and is a precondition for scaling the framework to networks beyond hand-curated topologies.

Fifth, all agents are backed by GPT-5.4~mini (\hyperref[appen:llm]{Section~\ref*{appen:llm}}); whether the reported MSR transfers to other model families is open.

Sixth, the single-agent baseline ablates the multi-agent architecture but does not calibrate the framework against expert manual work. 
A controlled study with security engineers producing mitigations under comparable constraints would situate the reported MSR relative to a human reference point.

%% file: Content/Conclusion.tex
\section{Conclusion}
\label{sec:conclusion}

We presented COHORT, the first end-to-end framework to autonomously generate and validate deployable mitigations against an observed adversary in emulated network. The two bottlenecks of expert dependency and production-exposure risk are addressed by a multi-agent \ac{LLM} generation pipeline and a high-fidelity emulator running real vendor firmware; each candidate is validated by offensive replay of the original Caldera adversary on the mitigated network. Across three topologies and four attack scenarios, COHORT outperforms a single-agent baseline (Section~\ref{sec:defense-success}).

The same autonomy opens new attack surfaces: a routine hardening run surfaced a long-acknowledged container-escape weakness, making staging isolation and human review of accepted mitigations preconditions for safe operation.

Results are scoped to scripted Caldera adversaries on a Linux-only lab with FortiGate, Cisco IOS, and Open vSwitch under GPT-5.4~mini, and connectivity preservation is verified by two probes; cross-attack transfer, adaptive adversaries, other model families, calibration against expert-derived mitigations, and finer operational regression remain open (Section~\ref{dis:open-questions}).

%% file: Content/Acknowledgements.tex
\section*{LLM usage}
Grammar was edited using Grammarly and Claude.

%% file: Content/Appendix.tex
\section{Appendix}

\subsection{Experimental Setup}

\subsubsection{Emulation Platform} \label{appen:gns3}
\href{https://www.gns3.com/}{GNS3} version 2.2.57 for our network emulation environment.

\subsubsection{Multi-Agent System} \label{appen:autogen}
\href{https://github.com/microsoft/autogen}{AutoGen version 0.7.5}~\cite{wu2024autogen, AutoGen}.

\subsubsection{Language Model} \label{appen:llm}
All agents were backed by OpenAI \texttt{GPT-5.4~mini}, snapshot version \texttt{2026-03-17}, served via an Azure OpenAI Global Standard deployment.

\subsubsection{Azure Server Specifications} \label{appen:azure}
Experiments ran on a \href{https://azure.microsoft.com/en-us/}{Microsoft Azure} Standard D8s v3 instance (8~vCPUs, 32~GiB RAM) running Ubuntu 24.04.2 LTS.

\subsubsection{Caldera} \label{appen:caldera}
\href{https://caldera.mitre.org/}{MITRE Caldera} version 5.3.0.

\subsubsection{Caldera Linux Agent} \label{appen:caldera_linux}
A custom Docker container on \href{https://hub.docker.com/r/gns3/ubuntu}{gns3/ubuntu:noble} running the \href{https://caldera.readthedocs.io/en/latest/plugins/sandcat/Sandcat-Details.html}{Caldera Sandcat client}. The agent reaches the Caldera C2 over HTTPS (port~8443) via a dedicated out-of-band management interface, so C2 connectivity survives mitigations applied to the topology interface.

\subsubsection{SSH Victim Container} \label{appen:ssh_victim}
A second container used as the lateral-movement target by the \textit{Lateral\_Movement} and \textit{Ransomware} adversaries. It runs OpenSSH on port~22 with password authentication for a non-root user with a deliberately weak password, and pre-populates the user's home directory with sensitive files for collection and encryption. The entry point emits periodic ICMP/HTTP background traffic to simulate an active host. No Caldera agent is installed.

\subsubsection{Firewall} \label{appen:firewall}
\href{https://www.gns3.com/marketplace/appliances/fortigate}{FortiGate} version 7.0.14.

\subsubsection{Switch}\label{appen:switch}
\href{https://gns3.com/marketplace/appliances/open-vswitch}{Open vSwitch} 3.3.7.

\subsubsection{Router}\label{appen:router}
\href{https://www.gns3.com/marketplace/appliances/cisco-3640}{Cisco 3640} running IOS 12.4(25d).

\subsection{Adversary Attack Steps} \label{appen:adversary-steps}
Table~\ref{tab:adv_all} enumerates, in execution order, the Caldera abilities that make up each adversary, together with their MITRE ATT\&CK tactic and technique identifiers. The final two steps in every adversary (LAN and internet connectivity checks) are shared validation abilities used to confirm that baseline network reachability is preserved after mitigation. The full ability definitions and downloadable Caldera adversary YAMLs are released alongside this paper as described in Section~\ref{appen:open-science}.

\input{Content/tables/pairwise_msr}

\input{Content/tables/cost_stats.tex}

\begin{figure}[pos=htbp]
    \centering
    \includegraphics[width=\columnwidth]{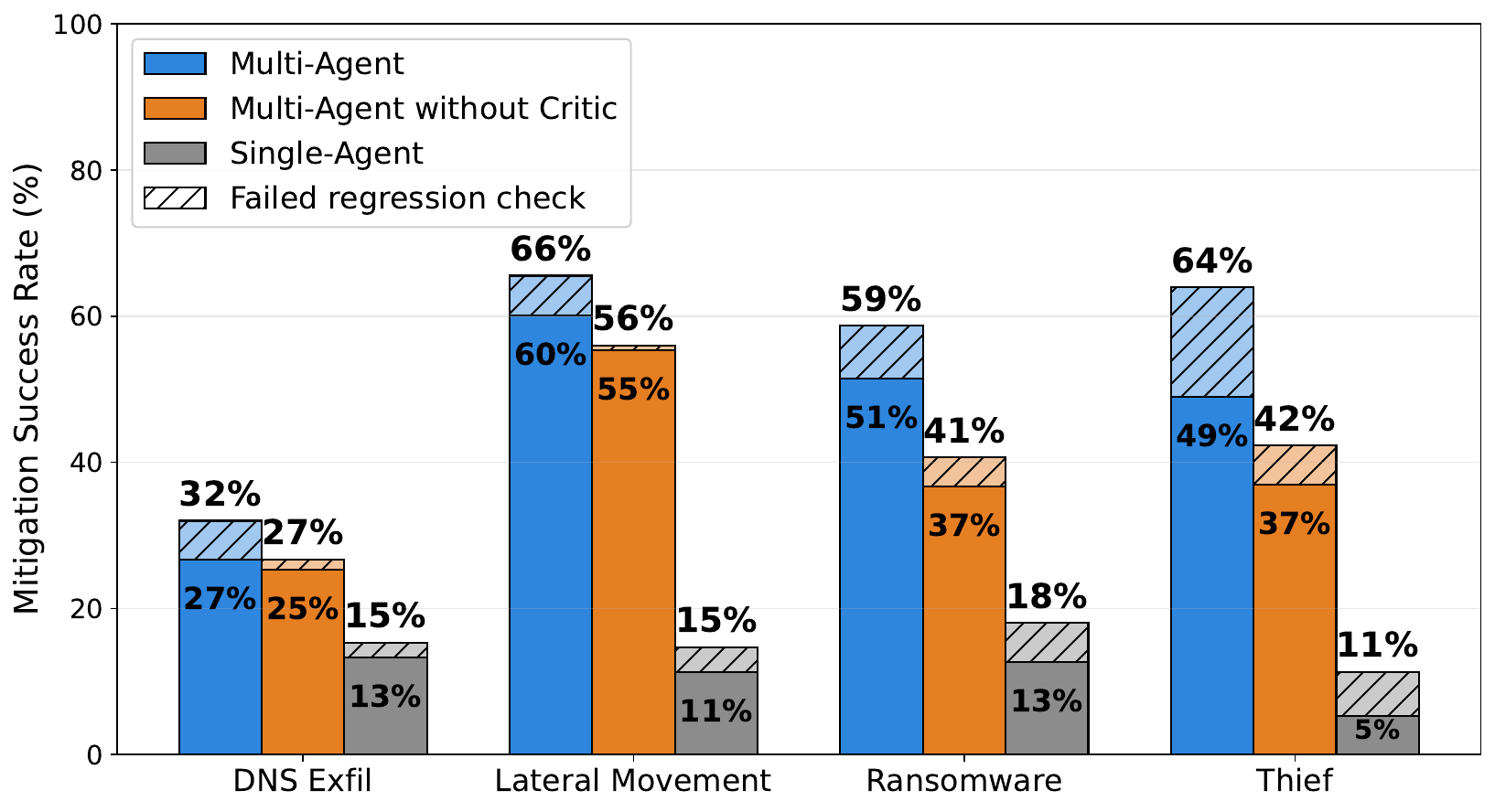}
    \caption{MSR by runtime condition and attack scenario. Hatched bar: rate without the connectivity-regression check.}
    \label{fig:msr_by_condition_attack}
\end{figure}

\input{Content/tables/adversaries_table}

\input{Content/Open_Science}

\subsection{Network Topologies}
\begin{figure}[pos=htbp]
    \centering
    \includegraphics[width=0.8\linewidth]{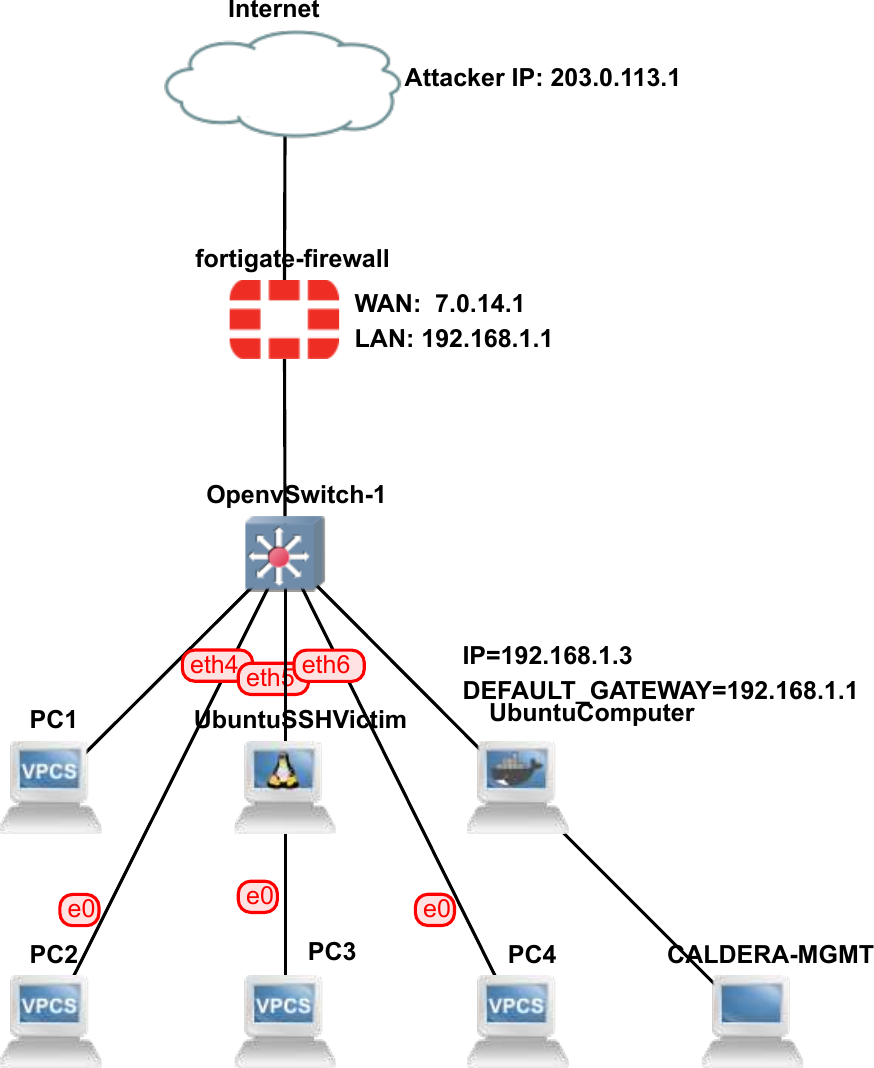}
    \caption{Small enterprise network topology.}
    \label{fig:small_enterprise}
    
\end{figure}

The medium enterprise network topology was inspired by \href{https://www.gns3.com/marketplace/labs/simple-network-layout}{Simple Network Layout} by Gilbert Nims (GNS3 Marketplace).
\begin{figure*}[pos=t]
    \centering
    \includegraphics[width=\textwidth]{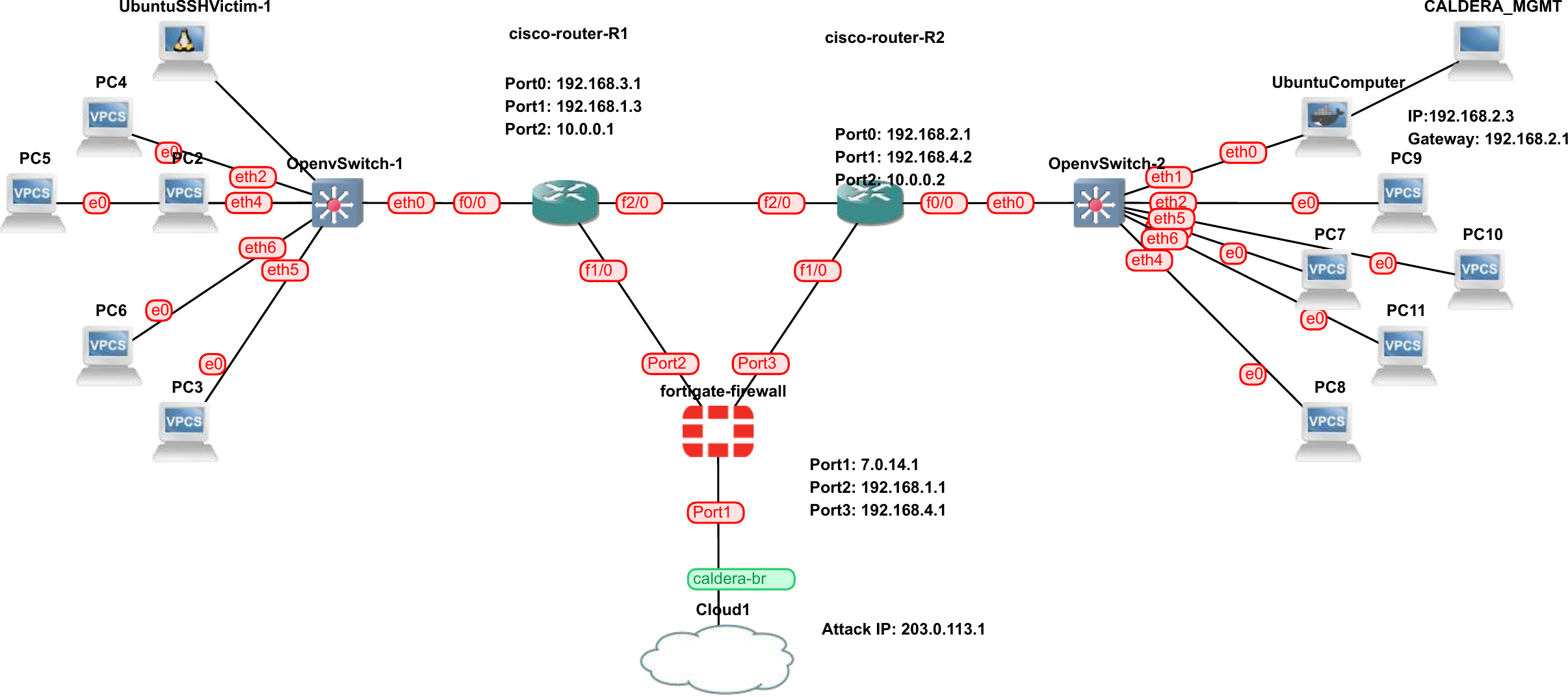}
    \caption{Medium enterprise network topology (inspired by \href{https://www.gns3.com/marketplace/labs/simple-network-layout}{Simple Network Layout} by Gilbert Nims, GNS3 Marketplace).}
    \label{fig:medium_enterprise}
\end{figure*}

The large enterprise network topology was inspired by \href{https://www.gns3.com/marketplace/labs/enterprise-network-lab-bank-project}{Enterprise Network Lab: Bank Project} by Kiki Oyewole (GNS3 Marketplace).
\begin{figure*}[pos=t]
    \centering
    \includegraphics[width=\textwidth]{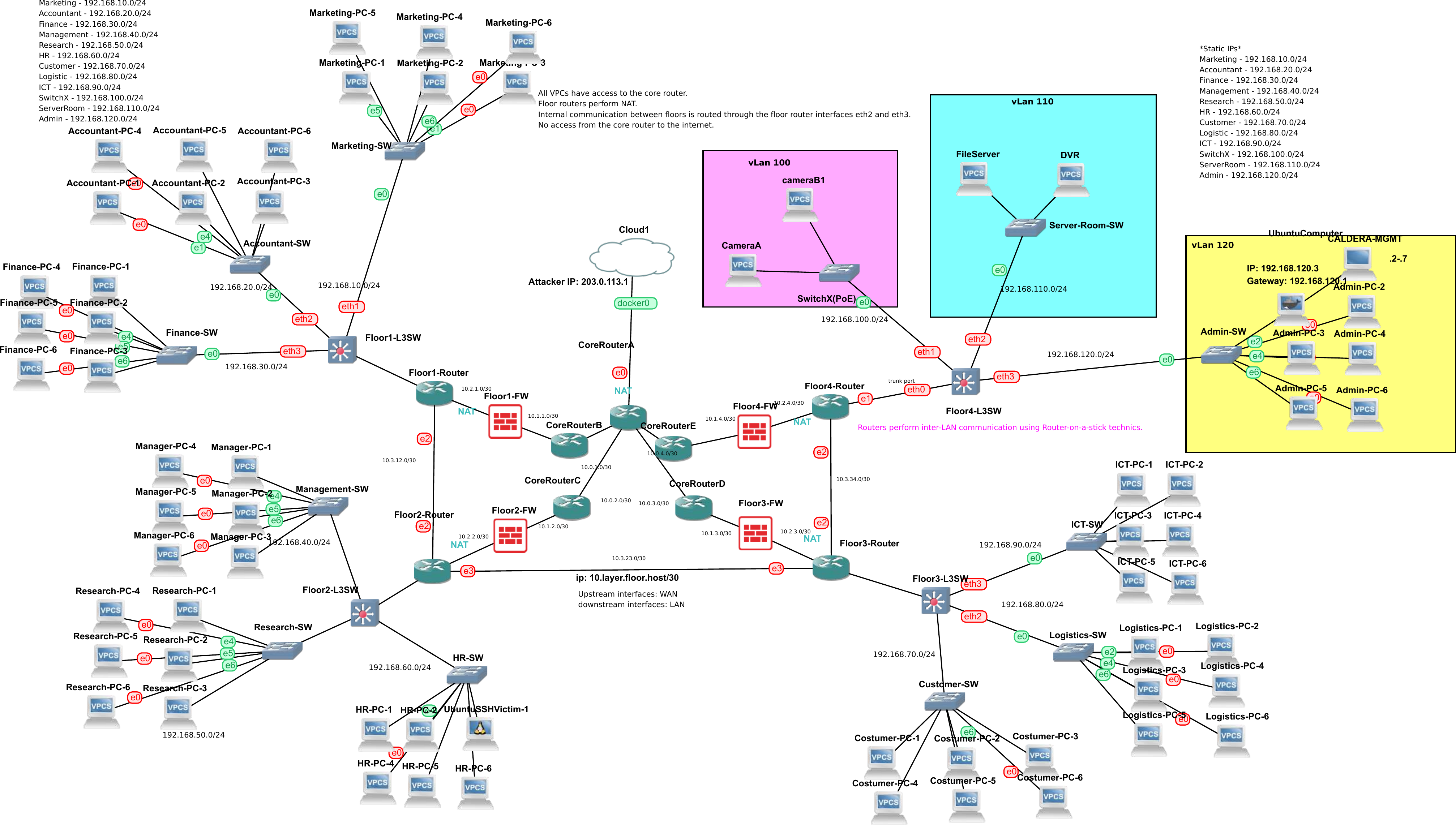}
    \caption{Large enterprise network topology (inspired by \href{https://www.gns3.com/marketplace/labs/enterprise-network-lab-bank-project}{Enterprise Network Lab: Bank Project} by Kiki Oyewole, GNS3 Marketplace).}
    \label{fig:large_enterprise}
\end{figure*}

\subsection{Pairwise Statistical Comparisons}\label{appen:pairwise-stats}

Table~\ref{tab:pairwise_stats} reports per-attack and overall pairwise condition comparisons referenced in Section~\ref{sec:defense-success}.
Each test is a Welch's t-test on run-level mean MSR (one observation per run; n=\nRunsSingle / \nRunsMultiNoCritic / \nRunsMultiCritic for Single-Agent / Multi-Agent without Critic / Multi-Agent), with Holm-Bonferroni correction across the three pairs.

\subsection{MSR by Attack Scenario}\label{appen:msr-by-attack}

Figure~\ref{fig:msr_by_condition_attack} disaggregates the headline MSR (Figure~\ref{fig:msr_by_condition}) by attack scenario. The hatched bar shows the rate without the connectivity-regression check. Sample sizes range from 138 to 150 mitigations per cell. Multi-Agent MSR ranges from $\msrAttackDNSExfilCritic\%$ on DNS Exfiltration to $\msrAttackLatMovCritic\%$ on Lateral Movement; DNS Exfiltration's lower rate is structural, reflecting that its egress channel (DNS) bypasses the network-based egress controls the suggester most often proposes.

\subsection{Per-Mitigation Cost}\label{appen:cost}

Table~\ref{tab:cost_stats} reports per-mitigation token usage, duration, and dollar cost by runtime condition. Each mitigation may comprise multiple implementation--critique iterations; the row aggregates across all such iterations within the mitigation. Cost is computed at GPT-5.4-mini Azure list pricing (\$0.75 per 1M input tokens, \$4.50 per 1M output tokens); the Total column sums across all mitigations in the condition.

%% file: Content/tables/pairwise_msr.tex
\begin{table}[t]
\centering
\small
\caption{Per-attack and overall pairwise condition comparisons on the strict joint MSR (attack disrupted \emph{and} connectivity preserved). Each test is a Welch's t-test on run-level mean MSR (one observation per run; $n=59$ / $n=60$ / $n=60$ for Multi-Agent / Multi-Agent without Critic / Single-Agent). Each cell shows the MSR difference ($\Delta$pp, improvement) and $p$-value, with Holm-Bonferroni correction across the three pairs ($\alpha=0.05$).}
\label{tab:pairwise_stats}
\setlength{\tabcolsep}{3pt}
\resizebox{\columnwidth}{!}{%
\begin{tabular}{lccccc}
\toprule
\textbf{Comparison} & \textbf{DNS Exfiltration} & \textbf{Lateral Movement} & \textbf{Ransomware} & \textbf{Thief} & \textbf{Overall} \\
\midrule
\shortstack[l]{Single-Agent $\to$\\ Multi-Agent without Critic} & \shortstack{+12.0 \\ ($p{=}0.038$)} & \shortstack{+44.0 \\ ($p{<}10^{-7}$)} & \shortstack{+24.0 \\ ($p{<}10^{-4}$)} & \shortstack{+31.6 \\ ($p{<}10^{-4}$)} & \shortstack{+27.9 \\ ($p{<}10^{-14}$)} \\
\shortstack[l]{Multi-Agent without Critic $\to$\\ Multi-Agent} & \shortstack{+1.3 \\ ($p{=}0.83$)} & \shortstack{+4.7 \\ ($p{=}0.42$)} & \shortstack{+14.4 \\ ($p{=}0.022$)} & \shortstack{+11.4 \\ ($p{=}0.13$)} & \shortstack{+7.9 \\ ($p{=}0.036$)} \\
\shortstack[l]{Single-Agent $\to$\\ Multi-Agent} & \shortstack{+13.3 \\ ($p{=}0.0025$)} & \shortstack{+48.7 \\ ($p{<}10^{-9}$)} & \shortstack{+38.4 \\ ($p{<}10^{-6}$)} & \shortstack{+43.0 \\ ($p{<}10^{-6}$)} & \shortstack{+35.8 \\ ($p{<}10^{-19}$)} \\
\bottomrule
\end{tabular}%
}
\end{table}

%% file: Content/tables/cost_stats.tex

\begin{table}[t]
\centering
\caption{Per-mitigation cost by runtime condition (mean $\pm$ std). Tokens are summed across all agent-role LLM calls within a mitigation (which may include multiple implementation--critique iterations); duration is the wall-clock window from the first to the last context file in the mitigation directory; cost applies GPT-5.4-mini Azure pricing of \$0.75 per 1M input tokens and \$4.50 per 1M output tokens.}
\label{tab:cost_stats}

\footnotesize
\renewcommand{\arraystretch}{1.2}
\setlength{\tabcolsep}{10pt}

\resizebox{\columnwidth}{!}{%
\begin{tabular}{lrrrrrr}
\toprule

\textbf{Condition}
&
\textbf{N}
&
\textbf{Input (k)}
&
\textbf{Output (k)}
&
\textbf{Duration (s)}
&
\textbf{Cost/mit (\$)}
&
\textbf{Total (\$)}
\\

\midrule

Multi-Agent
&
591
&
$609 \pm 546$
&
$3.8 \pm 1.4$
&
$124 \pm 50$
&
$0.47 \pm 0.41$
&
279.72
\\

\makecell[l]{Multi-Agent\\without Critic}
&
600
&
$273 \pm 195$
&
$2.6 \pm 0.4$
&
$96 \pm 45$
&
$0.22 \pm 0.15$
&
130.04
\\

Single-Agent
&
604
&
$109 \pm 77$
&
$1.8 \pm 0.3$
&
$60 \pm 18$
&
$0.09 \pm 0.06$
&
54.43
\\

\bottomrule
\end{tabular}%
}

\end{table}

%% file: Content/tables/adversaries_table.tex
\begin{table*}[!p]
\centering
\caption{Caldera adversary steps for the four post-compromise scenarios. The final two rows in every adversary (LAN/Internet connectivity checks) are shared validation abilities. The Thief adversary is adapted from the \textit{Thief} pack in the \href{https://github.com/mitre/stockpile/blob/master/data/adversaries/packs/1a98b8e6-18ce-4617-8cc5-e65a1a9d490e.yml}{MITRE Caldera Stockpile plugin}~\cite{Stockpile, caldera-thief-blog}; the two final connectivity-validation steps are our addition.}
\label{tab:adv_all}
\scriptsize

\begin{subtable}[t]{0.49\textwidth}
\centering
\caption{Lateral\_Movement adversary steps.}
\label{tab:adv_lateral_movement}
\begin{tabularx}{\linewidth}{@{}>{\raggedright\arraybackslash}p{0.3cm}>{\raggedright\arraybackslash}p{1.9cm}>{\raggedright\arraybackslash}p{1.3cm}>{\raggedright\arraybackslash}p{1.1cm}>{\raggedright\arraybackslash}X@{}}\toprule
\textbf{\#} & \textbf{Step} & \textbf{Tactic} & \textbf{Tech.\ ID} & \textbf{Description} \\
\midrule
1 & Scan target for open SSH port & Discovery & T1046 & nmap confirms port 22 is open on the lateral target \\
\midrule
2 & SSH password brute force & Credential Access & T1110.003 & Spray a short password list against SSH until a credential succeeds \\
\midrule
3 & SSH lateral movement & Lateral Movement & T1021.004 & Log into the remote host with the discovered credential \\
\midrule
4 & Remote host enumeration & Discovery & T1082 & Enumerate users, processes, and network config on the remote host \\
\midrule
5 & Collect files from remote host & Collection & T1005 & Locate sensitive files on the remote host and copy them back via SCP \\
\midrule
6 & Exfil remotely collected file to C2 & Exfiltration & T1041 & Upload the collected file to the Caldera C2 server \\
\midrule
7 & LAN Connectivity Check (Ping) & Discovery & T1018 & Validate that LAN reachability is preserved \\
\midrule
8 & Internet Connectivity Check (HTTP) & Discovery & T1016 & Validate that outbound internet reachability is preserved \\
\bottomrule
\end{tabularx}
\end{subtable}\hfill
\begin{subtable}[t]{0.49\textwidth}
\centering
\caption{Thief adversary steps.}
\label{tab:adv_thief}
\begin{tabularx}{\linewidth}{@{}>{\raggedright\arraybackslash}p{0.3cm}>{\raggedright\arraybackslash}p{1.9cm}>{\raggedright\arraybackslash}p{1.3cm}>{\raggedright\arraybackslash}p{1.1cm}>{\raggedright\arraybackslash}X@{}}\toprule
\textbf{\#} & \textbf{Step} & \textbf{Tactic} & \textbf{Tech.\ ID} & \textbf{Description} \\
\midrule
1 & Create staging directory & Collection & T1074.001 & Create a local directory to hold files for exfiltration \\
\midrule
2 & Find files & Collection & T1005 & Locate files deemed sensitive on the compromised host \\
\midrule
3 & Stage sensitive files & Collection & T1074.001 & Copy the discovered files into the staging directory \\
\midrule
4 & Compress staged directory & Exfiltration & T1560.001 & Archive the staging directory with a standard utility \\
\midrule
5 & Exfil staged directory & Exfiltration & T1041 & Upload the archive to the C2 server over the agent channel \\
\midrule
6 & LAN Connectivity Check (Ping) & Discovery & T1018 & Validate that LAN reachability is preserved \\
\midrule
7 & Internet Connectivity Check (HTTP) & Discovery & T1016 & Validate that outbound internet reachability is preserved \\
\bottomrule
\end{tabularx}
\end{subtable}

\vspace{2ex}

\begin{subtable}[t]{0.49\textwidth}
\centering
\caption{DNS\_Exfil adversary steps.}
\label{tab:adv_dns_exfil}
\begin{tabularx}{\linewidth}{@{}>{\raggedright\arraybackslash}p{0.3cm}>{\raggedright\arraybackslash}p{1.9cm}>{\raggedright\arraybackslash}p{1.3cm}>{\raggedright\arraybackslash}p{1.1cm}>{\raggedright\arraybackslash}X@{}}\toprule
\textbf{\#} & \textbf{Step} & \textbf{Tactic} & \textbf{Tech.\ ID} & \textbf{Description} \\
\midrule
1 & Find files & Collection & T1005 & Locate files deemed sensitive on the compromised host \\
\midrule
2 & Encode sensitive data for DNS exfiltration & Defense Evasion & T1027 & Base64-encode the contents of a discovered file into DNS-safe chunks \\
\midrule
3 & Exfiltrate data via DNS queries & Exfiltration & T1048.003 & Send encoded chunks as subdomain lookups to bypass HTTP egress controls \\
\midrule
4 & Remove DNS exfil artifacts & Defense Evasion & T1070.004 & Delete local staging files and clear shell history to cover tracks \\
\midrule
5 & LAN Connectivity Check (Ping) & Discovery & T1018 & Validate that LAN reachability is preserved \\
\midrule
6 & Internet Connectivity Check (HTTP) & Discovery & T1016 & Validate that outbound internet reachability is preserved \\
\bottomrule
\end{tabularx}
\end{subtable}\hfill
\begin{subtable}[t]{0.49\textwidth}
\centering
\caption{Ransomware adversary steps.}
\label{tab:adv_ransomware}
\begin{tabularx}{\linewidth}{@{}>{\raggedright\arraybackslash}p{0.3cm}>{\raggedright\arraybackslash}p{1.9cm}>{\raggedright\arraybackslash}p{1.3cm}>{\raggedright\arraybackslash}p{1.1cm}>{\raggedright\arraybackslash}X@{}}\toprule
\textbf{\#} & \textbf{Step} & \textbf{Tactic} & \textbf{Tech.\ ID} & \textbf{Description} \\
\midrule
1 & System Recon & Discovery & T1082 & Gather basic system information before ransomware deployment \\
\midrule
2 & Find files & Collection & T1005 & Locate files deemed sensitive for encryption \\
\midrule
3 & Encrypt sensitive files & Impact & T1486 & Generate a random AES key and encrypt the discovered files \\
\midrule
4 & Write ransom note & Impact & T1491.001 & Drop a ransom note on the victim host \\
\midrule
5 & Plant cron persistence & Persistence & T1053.003 & Install a cron job that beacons back to the C2 server \\
\midrule
6 & Exfil encryption key to C2 & Exfiltration & T1041 & Upload the AES encryption key to the C2 server over HTTP \\
\midrule
7 & LAN Connectivity Check (Ping) & Discovery & T1018 & Validate that LAN reachability is preserved \\
\midrule
8 & Internet Connectivity Check (HTTP) & Discovery & T1016 & Validate that outbound internet reachability is preserved \\
\bottomrule
\end{tabularx}
\end{subtable}
\end{table*}

%% file: Content/Open_Science.tex
\subsection{Data and Code Availability} \label{appen:open-science}

In the spirit of open science, we release the following artifacts at
\url{https://github.com/user32133/cohort} and
\url{https://cohort-experiments-app.streamlit.app/}:

\begin{itemize}
  \item Agent prompts for all roles (Suggester, Implementer, Critic,
    Judge, Summarizer, and the single-agent baseline).
  \item GNS3 topology files and device configurations for the small,
    medium, and large environments.
  \item Docker images for the containers used in the experiments,
    including images prepared but not used in the final evaluation.
  \item Caldera adversaries (Thief, Ransomware, DNS Exfiltration,
    Lateral Movement) and the abilities that compose them, with
    ATT\&CK mappings.
  \item Conversation logs and post-mitigation operational reports
    from every run analyzed in this work.
\end{itemize}

A demo video walking through the framework is available at \url{https://youtu.be/8Lj6rXAOwM8}.

The multi-agent runtime is \href{https://github.com/microsoft/autogen}{AutoGen} (Section~\ref{appen:autogen}). 
The agent code and the GNS3 automation code cannot be placed in the public domain; however, organizations involved in this research are open to reviewing disclosure requests submitted through the corresponding author.